\documentclass[%
reprint,superscriptaddress,
fnobibnotes,amsmath,amssymb,aps,
longbibliography,floatfix,]{revtex4-1}

\usepackage{graphicx}% Include figure files
\usepackage{dcolumn}% Align table columns on decimal point
\usepackage{bm}% bold math
\usepackage{svg}
\usepackage{xcolor}
\usepackage[colorlinks = true,
            linkcolor = blue,
            urlcolor  = blue,
            citecolor = magenta,
            anchorcolor = blue]{hyperref}
\usepackage{enumitem}
\usepackage{siunitx}

\usepackage{amsmath} % or simply amstext
\newcommand{\angstrom}{\text{\normalfont\AA}}

\newcommand{\eref}[1]{Eq.~(\ref{#1})}

\newcommand{\ocite}[1]{Ref.~\cite{#1}}
\newcommand{\ra}{\rangle}
\newcommand{\la}{\langle}
\newcommand{\tot}{\langle \tilde{O}\rangle_t}
\newcommand{\ton}{\langle \tilde{O}\rangle_{ncc}}
\newcommand{\lam}{\lambda}
\makeatletter
\renewcommand{\fnum@figure}{\textbf{Fig.~\thefigure}}
\makeatother

\begin{document}

\title{Noise-Robust Estimation of Quantum Observables in Noisy Hardware}

\author{Amin Hosseinkhani}
\email{amin.hosseinkhani@meetiqm.com}
\affiliation{IQM Quantum Computers, Georg-Brauchle-Ring 23-25, 80992 Munich, Germany}
\author{Fedor \v{S}imkovic}
\affiliation{IQM Quantum Computers, Georg-Brauchle-Ring 23-25, 80992 Munich, Germany}
\author{Alessio Calzona}
\affiliation{IQM Quantum Computers, Georg-Brauchle-Ring 23-25, 80992 Munich, Germany}
\author{Emiliano Godinez-Ramirez}
\affiliation{IQM Quantum Computers, Georg-Brauchle-Ring 23-25, 80992 Munich, Germany}
\author{Vicente Pina-Canelles}
\affiliation{IQM Quantum Computers, Georg-Brauchle-Ring 23-25, 80992 Munich, Germany}
\affiliation{Department of Physics and Arnold Sommerfeld Center for Theoretical Physics, Ludwig-Maximilians-Universität München, Theresienstr. 37, 80333 Munich, Germany}
\author{Tianhan Liu}
\affiliation{IQM Quantum Computers, Georg-Brauchle-Ring 23-25, 80992 Munich, Germany}
\author{José D. Guimarães}
\affiliation{IQM Quantum Computers, Georg-Brauchle-Ring 23-25, 80992 Munich, Germany}
\author{Adrian Auer}
\affiliation{IQM Quantum Computers, Georg-Brauchle-Ring 23-25, 80992 Munich, Germany}
\author{In\'es de Vega}
\affiliation{IQM Quantum Computers, Georg-Brauchle-Ring 23-25, 80992 Munich, Germany}
\affiliation{Department of Physics and Arnold Sommerfeld Center for Theoretical Physics, Ludwig-Maximilians-Universität München, Theresienstr. 37, 80333 Munich, Germany}

\date{\today}
\begin{abstract}
Error mitigation is essential for extracting reliable results from quantum computations performed on noisy intermediate-scale quantum hardware. Here we introduce Noise-Robust Estimation (NRE), a noise-agnostic framework that suppresses estimation bias through a two-stage post-processing protocol. The method combines measurement data from a target circuit and a corresponding noise-canceling companion circuit to construct a baseline estimator with reduced sensitivity to noise. We show that the residual bias of this estimator is governed by the variation of an auxiliary quantity across amplified noise realizations, motivating the use of a measurable diagnostic quantity—the normalized dispersion of this auxiliary estimator. When the dispersion approaches zero, contributions arising from imperfect noise amplification vanish and the remaining bias terms are expected to diminish for smooth stationary noise profiles. Leveraging this relationship, NRE performs a final extrapolation to the zero-dispersion limit using bootstrapped measurement data. We experimentally validate the method on a 20-qubit IQM superconducting quantum processor using circuits containing up to 480 entangling CZ gates. Across a variety of circuits and noise levels, NRE consistently achieves substantially reduced bias compared to existing mitigation techniques while maintaining moderate sampling overhead. These results establish NRE as a practical and broadly applicable error-mitigation strategy for quantum computations on noisy hardware.
\end{abstract}

\maketitle
\section{\label{sec:intro}Introduction}

The practical utility of quantum computing depends not only on increasing qubit counts but also on mitigating the noise that disrupts quantum computations. While fully fault-tolerant quantum computing remains a long-term goal, quantum error mitigation (QEM) is essential for obtaining reliable results from today’s Noisy Intermediate-Scale Quantum (NISQ) devices and for improving the performance of partially error-corrected quantum processors. Even as quantum error correction (QEC) advances, QEM will remain relevant for suppressing residual errors and extending the effective fidelity of quantum computations in regimes where full fault tolerance is impractical. Numerous mitigation techniques have been proposed for estimating expectation values~\cite{Endo_2021, Cai_2023}. However, existing approaches incur sampling overheads that grow exponentially with circuit size and per-qubit error rates~\cite{Takagi_2023, Tsubouchi_2023, Quek_2024, Filippov_2024}, making scalable error mitigation a central challenge.

Broadly, quantum error mitigation methods can be categorized as either noise-aware or noise-agnostic. Noise-aware approaches, such as probabilistic error cancellation (PEC)~\cite{Temme_2017, Endo_2018, Song_2019, Ewout_2023}, rely on detailed noise models to reconstruct ideal results. While PEC can in principle eliminate the impact of noise, its effectiveness depends critically on accurate noise characterization. Fundamental limitations in the learnability of Pauli noise~\cite{Chen_2023, Alessio_2024}, together with parameter drifts and model inaccuracies, can lead to finite residual bias~\cite{Govia_2024, Alessio_2024}.

\begin{figure*}
    \centering
    \includegraphics[width=0.7\linewidth]{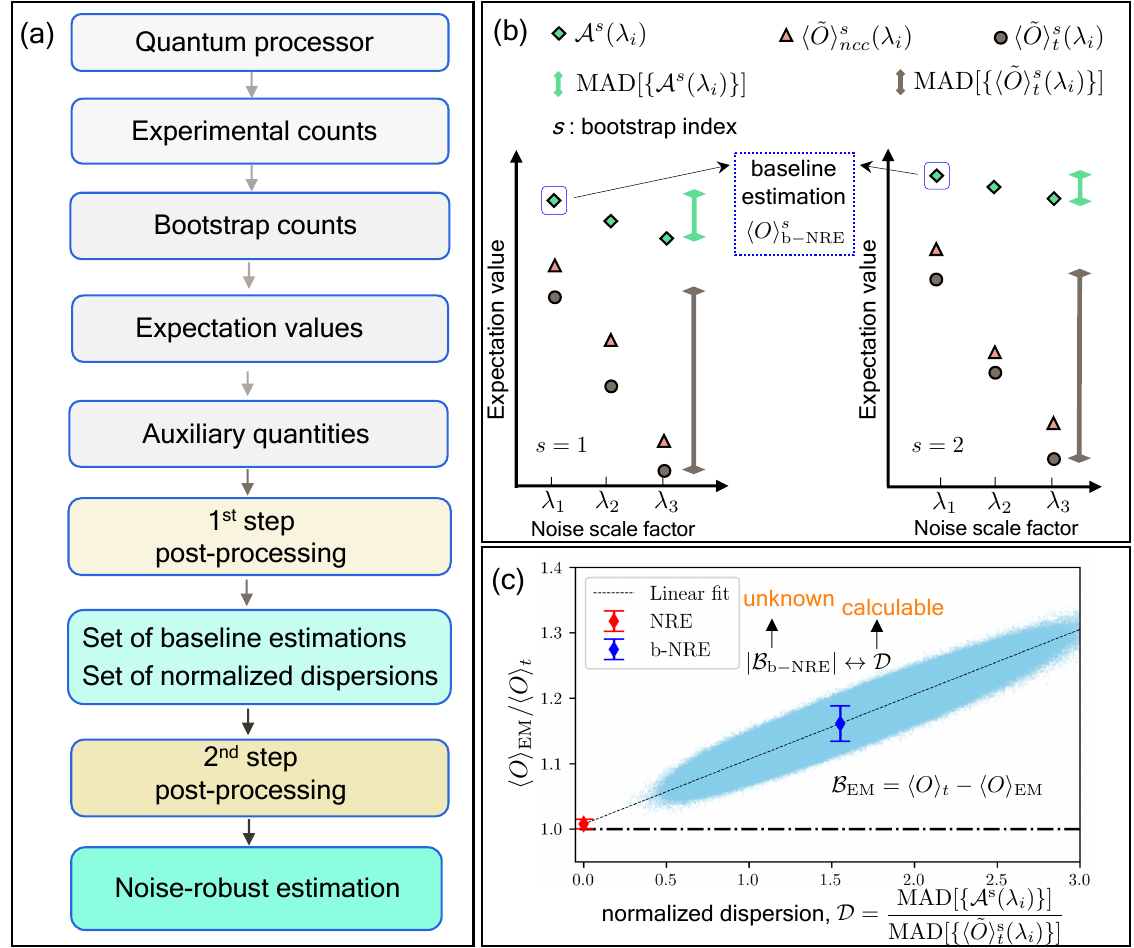}
\caption{(a) Workflow of NRE, with each step detailed in Sec.~\ref{sec:NRE_workflow}.
(b) Schematic illustration of expectation values obtained from executing the target circuit $\langle \tilde{O} \rangle_t$, the noise-canceling circuit $\langle \tilde{O} \rangle_{ncc}$, and the auxiliary quantity $\mathcal{A}$ as a function of the noise scale factor $\lambda_i$. Expectation values and $\mathcal{A}$ are shown for two different bootstrap indices ($s$). The mean absolute deviation (MAD) of the sets $\{\langle \tilde{O}^s_t(\lambda_i)\rangle\}$ and $\{\mathcal{A}^s(\lambda_i)\}$ is schematically represented. As explained in Sec.~\ref{sec:NRE_workflow}, the first post-processing step ensures that $\mathcal{A}(\lambda_1)$ serves as a baseline estimator for the ideal expectation value.
(c) Experimentally observed correlation between the normalized dispersion
$\mathcal{D}$ and the residual bias of the baseline estimator,
$\mathcal{B}_{\mathrm{b\text{-}NRE}}$. The light blue cloud represents the distribution of baseline estimations obtained from different bootstrap realizations. Blue and red points correspond to the mean and standard deviation of the baseline estimator (b-NRE) and the final estimator (NRE), respectively. The residual bias of an error-mitigated estimator is defined as
$\mathcal{B}_{\mathrm{EM}} = \langle O\rangle_{t}-\langle O\rangle_{\mathrm{EM}}$,
where $\langle O\rangle_{t}$ denotes the ideal expectation value of the target circuit and $\langle O\rangle_{\mathrm{EM}}$ denotes the expectation value obtained from an error-mitigation estimator. While the bias can take both positive and negative values, the magnitude of the baseline-estimator bias, $|\mathcal{B}_{\mathrm{b\text{-}NRE}}|$, is expected to correlate with the normalized dispersion $\mathcal{D}$ under mild smoothness assumptions on the noise response, enabling extrapolation toward the $\mathcal{D}\to 0$ limit. A detailed explanation of the experiment is provided in Sec.~\ref{sec:TFIM_implementation}.
}
    \label{fig:NRE_Schematics}
\end{figure*}

Noise-agnostic approaches avoid explicit noise modeling. A prominent example is zero-noise extrapolation (ZNE)~\cite{Temme_2017, Li_2017, IBM_utility2023}, which estimates noiseless expectation values by amplifying noise and extrapolating to the zero-noise limit. However, ZNE faces two major challenges. First, in non-Clifford circuits subject to Pauli noise, expectation values often follow a multi-exponential decay~\cite{IBM_utility2023, Cai_2021} with an unknown number of components. As a result, commonly used extrapolation models—such as polynomial~\cite{Temme_2017, Li_2017, Alessio_2022} or single-exponential fits~\cite{Endo_2018, IBM_utility2023}—can introduce bias due to model mismatch. Second, noise amplification itself is often imperfect. Techniques such as pulse-level stretching~\cite{Temme_2017, IBM_ExpZNE2019} or gate-level unitary folding~\cite{Giurgica_Tiron_2020, Dumitrescu2018, He_PRA_2020} typically do not scale noise uniformly because experimental noise channels generally do not commute with circuit unitaries~\cite{koenig2024invertedcircuit}. Although probabilistic noise amplification~\cite{IBM_utility2023, Ferracin_2022, Hour2024} can improve scaling accuracy, it reintroduces dependence on noise modeling, undermining the noise-agnostic premise.

Other noise-agnostic strategies attempt to learn the effect of noise using auxiliary circuits. Clifford Data Regression (CDR)~\cite{Czarnik_2021} and its extension variable-noise CDR (vnCDR)~\cite{Lowe2021} use near-Clifford circuits to train regression models that correct noisy expectation values. While effective for shallow circuits, their accuracy typically degrades for deeper circuits due to mismatches in noise scaling. Similarly, the depolarizing noise mitigation method proposed by Urbanek \textit{et al.}~\cite{Urbanek_2021} mitigates local depolarizing noise by inserting estimation circuits prior to performing ZNE. However, this approach relies on the assumption that depolarizing noise dominates, limiting its accuracy on real devices with more complex error processes.

These limitations motivate the development of noise-agnostic mitigation methods that systematically suppress bias while maintaining manageable sampling costs. In this work we introduce Noise-Robust Estimation (NRE), a noise-agnostic framework for expectation-value estimation that combines noise amplification with a structured post-processing strategy. The central idea is to construct an auxiliary estimator whose variation across amplified noise realizations provides a directly measurable diagnostic of the magnitude of the residual bias of the baseline estimator.

NRE introduces an auxiliary quantity $\mathcal{A}$ that is designed to be less sensitive to noise than the directly measured expectation value $\tot$ of an observable $O$. The framework consists of two post-processing steps.

\textbf{First post-processing step: baseline estimation.}
An initial estimator, $\langle O \rangle_\mathrm{b\text{-}NRE}$, is constructed using measurements from both the target circuit and a noise-canceling circuit (\textit{ncc}) together with a tunable control parameter. The noise-canceling circuit is structurally similar to the target circuit but chosen such that its noiseless expectation value is known. By combining data from these circuits, this step suppresses leading noise contributions and produces a baseline estimator whose remaining bias arises from higher-order and finite-difference effects.

\textbf{Second post-processing step: bias--dispersion correlation.}
The second stage exploits the relationship between the magnitude of the residual bias of the baseline estimator, $|\mathcal{B}_{\mathrm{b\text{-}NRE}}|$, and the normalized dispersion $\mathcal{D}$ of the auxiliary estimator across amplified noise realizations. As shown in Appendix~\ref{sec:ap_residual_bias}, the residual bias can be decomposed into contributions arising from truncation error, discretization error, and imperfect noise amplification. The contribution associated with imperfect noise amplification vanishes when the auxiliary estimator becomes independent of the noise scale, corresponding to the limit $\mathcal{D}\to 0$. For smooth noise responses, the same derivative hierarchy that governs the residual-bias magnitude also controls the variation of the auxiliary estimator across noise scales (see Appendix~\ref{sec:ap_residual_bias}). Consequently, reduced variation of the auxiliary estimator across noise realizations is associated with a systematic suppression of the residual-bias magnitude. Motivated by this analytical structure, the final NRE estimate is obtained by extrapolating the baseline estimator to the $\mathcal{D}\to 0$ limit.

A distinctive feature of NRE is its use of classical bootstrapping, which serves two purposes: estimating statistical uncertainties and generating an ensemble of baseline estimators with different dispersion values. This bootstrap ensemble enables a regression procedure that extrapolates the estimator to the $\mathcal{D}\to 0$ limit, thereby suppressing residual bias.

Figure~\ref{fig:NRE_Schematics}(a) illustrates the NRE workflow, with each step detailed in Sec.~\ref{sec:NRE_workflow}. Figure~\ref{fig:NRE_Schematics}(b) schematically shows expectation values obtained from executing the target and noise-canceling circuits at multiple noise scale factors. Bootstrap realizations generate independent sets of expectation values and auxiliary estimators, whose dispersions are illustrated schematically. Figure~\ref{fig:NRE_Schematics}(c) shows the experimentally observed relationship between normalized dispersion and the residual bias of the baseline estimator.

Using global unitary folding to amplify noise, we experimentally validate the NRE framework on IQM Garnet, a 20-qubit superconducting quantum processor with a square topology~\cite{Garnet}. In Sec.~\ref{sec:TFIM_implementation}, we apply NRE to the measured ground-state energy of the transverse-field Ising model (TFIM), demonstrating substantial bias reduction compared with ZNE, CDR, vnCDR, and the method of Urbanek \textit{et al.} across a range of noise amplification settings.

In Sec.~\ref{sec:H4}, we apply NRE to a quantum chemistry problem by estimating the ground-state energy of the $\mathrm{H}_4$ molecule. Despite strong noise effects arising from deep circuits and high-weight observables, NRE restores the estimated energy close to the ideal value. Finally, in Sec.~\ref{subsec:SO}, we analyze the sampling overhead of NRE via numerical simulations and compare it with ZNE and the method proposed by Urbanek \textit{et al.}. Consistent with the experimental observations, exploiting the $|\mathcal{B}_{\mathrm{b\text{-}NRE}}|$–$\mathcal{D}$ relationship enables improved estimator accuracy while maintaining moderate sampling overhead.

Taken together, these results demonstrate that NRE provides a practical noise-agnostic error mitigation framework that overcomes several limitations of existing techniques. Unlike standard ZNE, NRE tolerates inaccuracies in noise scaling. Unlike CDR and vnCDR, which require large sets of training circuits, NRE requires only a single noise-canceling circuit per target circuit. Furthermore, extrapolation to the $\mathcal{D}\to0$ limit reduces the effect of mismatches between the noise responses of the target and noise-canceling circuits. More broadly, the connection between normalized dispersion and residual-bias magnitude uncovered in this work suggests that statistical diagnostics derived from measurement data can guide error mitigation strategies on noisy quantum devices.

We conclude in Sec.~\ref{sec:conclusion} by outlining potential future directions, including integration of NRE with complementary noise-suppression techniques and its application to early fault-tolerant quantum processors, where logical qubits are expected to retain residual noise and accurate estimation of logical observables will remain essential.

The appendices provide the analytical and implementation details underlying the NRE protocol. In particular, Appendix~\ref{sec:ap_residual_bias} analyzes the structure of the residual bias and its connection to the normalized dispersion, Appendix~\ref{subsec:n_op} derives the optimal control parameter entering the baseline estimator, Appendix~\ref{subsec:extended_NRE} describes the extended resampling procedure used to propagate statistical uncertainties, Appendix~\ref{sec:ap_ncc} details the construction of the noise-canceling circuits, and Appendix~\ref{sec:ap_regression_strategies} examines the robustness of the final NRE estimate with respect to the regression weighting strategy.
\section{NRE framework}
\label{sec:NRE_workflow}

The NRE framework estimates the noiseless expectation value of an observable by combining measurements from two circuits with similar structures but generally different noise responses. The key idea is to construct an auxiliary estimator whose noise sensitivity is reduced relative to the directly measured observable. This auxiliary quantity is evaluated across several amplified noise realizations, enabling the extraction of an accurate estimate of the ideal expectation value through a structured post-processing procedure. In the following, we formalize this procedure and introduce the quantities required for its implementation.

Let us assume that the physical noise strength affecting the circuit can be characterized by a parameter $\epsilon_0$. When noise amplification is applied, the effective noise strength becomes $\epsilon = \lambda \epsilon_0$, where $\lambda$ is a dimensionless noise scale factor. Our objective is to estimate the noiseless expectation value of an observable $O$ defined by the target quantum circuit, denoted by $\langle O\rangle_t$. Due to noise in the prepared quantum state and subsequent operations, the experimentally measured expectation value becomes a noise-dependent quantity, which we denote by $\tot(\lambda)$.

To perform NRE, we measure the observable $O$ using the target circuit. As in ZNE, measurements are performed at $M$ different noise scale factors $\lambda_i$, where $i=1,2,\dots,M$. Without loss of generality, we assume a uniform spacing of the noise levels, $\lambda_{i+1}-\lambda_i = h$. In addition to the target circuit, NRE requires measurements of the observable $O$ using a second circuit that we refer to as the \textit{noise-canceling circuit} (\textit{ncc}). Measurements of this circuit are performed at the same noise scale factors $\lambda_i$. The purpose of introducing the noise-canceling circuit is to provide a reference observable whose noiseless expectation value is known while retaining a noise response similar to that of the target circuit.

A key requirement for the noise-canceling circuit is that its noiseless expectation value, $\la O\ra_{ncc}$, must be known in advance. In this work, we achieve this by replacing all non-Clifford gates in the target circuit with Clifford gates, allowing the noiseless expectation value to be efficiently computed via classical simulation. The noise-canceling circuit thus retains the same gate structure and number of qubit operations as the target circuit. %Details of noise-canceling circuit generation from a given target circuit is presented in Appendix. 

To construct an estimator with reduced noise sensitivity, we combine measurements from the target and noise-canceling circuits in a manner that partially cancels their noise responses while preserving the ideal expectation value. The goal is to construct a quantity whose noiseless limit reproduces the target expectation value while reducing sensitivity to multiplicative noise distortions through the use of the noise-canceling circuit. We therefore introduce the NRE \textit{ansatz} in terms of an auxiliary quantity $\mathcal{A}$ evaluated at noise scale $\lambda_i$,

\begin{align}\label{eq:aux_O}
    &\mathcal{A}_{ncc}(n, \lambda_i) = \mathcal{P}_1 (\lambda_i)+ n\mathcal{P}_2 ( \lambda_i)
\end{align}

in which,

\begin{align}\label{eq:aux_P1}
    &\mathcal{P}_1 (\lambda_i) = \tot(\lambda_i)\left(\frac{\la O\ra_{ncc}}{\ton(\lambda_i)}\right),\\\label{eq:aux_P2}
    &\mathcal{P}_2 (\lambda_i) = \log\left(\frac{\la O\ra_{ncc}}{\ton(\lambda_i)}\right).  
\end{align}

Here the real number $n$ is the control parameter used for the first step of the post-processing. From the above definitions, it is clear that in the asymptotic limit where noise vanishes we have $\mathcal{P}_1\rightarrow \langle O\rangle_t$ and $\mathcal{P}_2\rightarrow 0$, so that

\begin{align}
\label{eq:lim}
    \lim_{\lambda \rightarrow 0} \mathcal{A}_{ncc}(n, \lambda) = \langle O\rangle_t.
\end{align}

The term $\mathcal{P}_1$ combines the measured expectation value of the target circuit with the ratio between the known noiseless expectation value of the noise-canceling circuit and its measured noisy value. This construction partially compensates for noise effects that are shared between the two circuits. Because the target and noise-canceling circuits have similar gate structures but generally different noise responses, the ratio appearing in $\mathcal{P}_1$ partially cancels shared noise contributions while remaining sensitive to imperfect matching between the two circuits. Consequently, one can generally expect $\mathcal{P}_1(\lambda_i)$ to exhibit reduced noise sensitivity compared with $\tot(\lambda_i)$. In many practical situations this behavior leads to a slower variation of $\mathcal{P}_1$ with increasing noise scale factor, which can locally be approximated by a linear function in the vicinity of the smallest noise scale factor ($\lambda_1=1$).

To further reduce the noise sensitivity of the auxiliary estimator, we introduce the term $\mathcal{P}_2$, defined in logarithmic form. The logarithm captures multiplicative deviations between the noisy and noiseless expectation values of the noise-canceling circuit. Because small multiplicative errors become approximately additive under the logarithm, $\mathcal{P}_2$ varies approximately linearly with the noise scale factor in the small-noise regime. This property allows the tunable control parameter $n$ to optimally adjust $\mathcal{A}$, enhancing its robustness against noise variations and enabling a more precise expectation value estimation.

In writing \eref{eq:aux_O}, we assume that for the entire range of considered noise scale factors (including $\lambda = 0$ for the \textit{ncc}), the sign of $\ton(\lambda_i)$ remains constant, ensuring that $\mathcal{P}_2$ is a real number. The validity of this assumption can be verified directly from the measured data.

Using Eq.~(\ref{eq:lim}), the noiseless expectation value can be written as a Taylor expansion of $\mathcal{A}$ around the smallest noise scale factor $\lambda_1$,

\begin{align}
\label{eq:Taylor_sum}
    \langle O\rangle_t = \mathcal{A}_{ncc}(n, \lambda_1) &+ \sum_{j=1}^{M-1}\frac{(-\lambda_1)^{j}}{j!}\mathcal{A}^{[j]}_{\lam_1}(n) \nonumber\\
    & + \mathcal{B}(M, n)
\end{align}

Here $\mathcal{A}^{[j]}_{\lam_1}(n)$ denotes the numerically approximated derivative of $\mathcal{A}$ of order $j$ taken at $\lambda_1$.

The term $\mathcal{B}$ represents the unknown residual bias  that arises when approximating $\la O\ra_t$ via a truncated Taylor expansion. This error originates from both truncation of the expansion and discretization errors in the numerical derivative evaluation. In addition, noise-scaling mismatch between $\tot(\lambda_i)$ and $\ton(\lambda_i)$ can affect the magnitude of these terms. Imperfect noise amplification may introduce further contributions to the residual bias through inaccuracies in the numerical derivative coefficients induced by mismatch between the intended and implemented noise scale factors. The general structure and properties of $\mathcal{B}(M,n)$ are discussed in detail in Appendix~\ref{sec:ap_residual_bias}.

The reduced noise sensitivity of $\mathcal{A}$ motivates estimating $\langle O\rangle_t$ through a Taylor expansion of $\mathcal{A}(\lambda)$ rather than $\tot(\lambda)$, as is done in conventional ZNE approaches. Because $\mathcal{A}$ is constructed to vary more slowly with the noise scale factor, the resulting truncation error—associated with higher-order derivatives—is expected to be smaller.

In the following, we describe the two post-processing steps used in the NRE framework.

\subsection{First post-processing step: Baseline estimation}

We define the optimal control parameter as the value that ensures the auxiliary quantity at the smallest noise level ($\lambda_1$) serves as an estimator for the ideal expectation value. The optimal control parameter $n_{op}$ therefore satisfies

\begin{align}
    \sum_{j=1}^{M-1}\frac{(-\lambda_1)^{j}}{j!}\mathcal{A}^{[j]}_{\lam_1}(n_{op}) = 0.
\end{align}

The solution to this equation is discussed in Appendix~\ref{subsec:n_op}. Using this value we obtain,

\begin{align}
\label{eq:NRE_estimation}
    \langle O \rangle_\mathrm{b-NRE} = \mathcal{A}_{ncc}(n_{op}, \lambda_1),
\end{align}
which serves as the baseline estimator for the ideal expectation value according to \eref{eq:Taylor_sum}. For the baseline estimator obtained at $n=n_{\mathrm{op}}$, we denote the corresponding residual bias by $\mathcal{B}_{\mathrm{b\text{-}NRE}}$.

Neglecting statistical uncertainties (shot noise), we note that in the ideal case where noise has no effect on $\mathcal{P}_1$ (i.e., $\mathcal{P}_1$ remains constant for all $\lambda_i$), the optimal control parameter simplifies to $n_{op}=0$. This ideal case occurs only when the noise-scaling properties of the target and noise-canceling circuits are perfectly matched. In general this condition is not fulfilled. Consequently, the baseline estimation depends on the specific choice of noise-canceling circuit.

In the specific case of $M=2$, setting the control parameter to its optimal value ensures that $\mathcal{A}_{ncc}(n_{op}, \lambda_1) = \mathcal{A}_{ncc}(n_{op}, \lambda_2)$. For $M\geq 3$, however, $\mathcal{A}_{ncc}(n_{op},\lambda_i)$ generally varies across different $\lambda_i$ due to residual noise effects. To further enhance the accuracy of the estimation for a fixed choice of the noise-canceling circuit, we therefore introduce a second post-processing step, which requires at least $M=3$ noise scale factors.

\subsection{Second post-processing step: Bias–dispersion correlation}
The residual bias  $\mathcal{B}_{\mathrm{b\text{-}NRE}}$ in the baseline estimation depends on higher-order derivatives of $\mathcal{A}$ at $\lambda_1$. Although these derivatives cannot be directly computed from the experimental data, their magnitude is reflected in the variation of $\mathcal{A}_{ncc}(n_{op}, \lambda_i)$ across the measured noise scale factors. In particular, a smaller variation of $\mathcal{A}_{ncc}(n_{op}, \lambda_i)$ across the sampled noise scales $\lambda_i$ indicates that the higher-order derivatives are reduced in magnitude, and therefore suggests that the truncation- and discretization-related components of the residual-bias magnitude are also smaller. Moreover, as discussed in Appendix~\ref{sec:ap_residual_bias}, modeling the effects of imperfect noise amplification shows that the corresponding contribution to $\mathcal{B}_{\mathrm{b\text{-}NRE}}$ also decreases as the variation of $\mathcal{A}_{ncc}(n_{op}, \lambda_i)$ becomes smaller. For smooth and approximately stationary noise profiles, reduced variation across noise scales is therefore expected to suppress multiple components of the residual bias simultaneously, which is reflected in a smaller normalized dispersion $\mathcal{D}$.

We therefore introduce the normalized dispersion $\mathcal{D}$, a dimensionless metric that quantifies the variation of the auxiliary estimator across noise scales and serves as a measurable diagnostic for the magnitude of the residual bias of the baseline estimator,

\begin{align}
    \mathcal{D} = \frac{\mathrm{MAD}[\{\mathcal{A}(\lambda_i)\}]}{\mathrm{MAD}[\{\tot(\lambda_i)\}]}.
    \label{eq:nd}
\end{align}

The mean absolute deviation of a dataset is defined as $\mathrm{MAD}[\{x_j\}] = (1/m)\sum_{j=1}^{m}|x_j -\bar{x}|$, where $\bar{x}$ denotes the mean of the set $\{x_j\}$.

To implement this step, we first apply classical bootstrapping to the experimental counts, generating an ensemble of bootstrapped datasets labeled by the index $s$. From each dataset we compute the corresponding expectation values $\tot^s(\lambda_i)$ and $\ton^s(\lambda_i)$.

For each bootstrap index $s$ we then evaluate the auxiliary quantity $\mathcal{A}_{ncc}^s$ and apply the first post-processing step to obtain the corresponding baseline estimator $\langle O\rangle_\mathrm{b-NRE}^s$. At the same time we compute the normalized dispersion $\mathcal{D}^s$. This procedure generates a dataset consisting of baseline estimators and their associated dispersion values.

In practice, the final NRE estimate is obtained by performing a regression on this dataset and extrapolating to the $\mathcal{D}^s\rightarrow 0$ limit,

\begin{align}
\langle O\rangle_\mathrm{NRE}\equiv \lim_{\mathcal{D}^s\rightarrow 0} {\langle O\rangle_\mathrm{b-NRE}^s}(\mathcal{D}^s).
\end{align}

Fig.~\ref{fig:NRE_Schematics}(a) illustrates the workflow of the NRE framework. This procedure yields a single final estimate. To determine the statistical uncertainty of the NRE estimator one may repeat the full procedure from the bootstrapping stage onward. In our implementation we employ a computationally more efficient approach in which additional samples are drawn from a normal distribution centered at each bootstrapped estimator, with variance determined from the full bootstrap ensemble. This procedure produces a distribution of NRE estimators whose size equals the number of bootstrap samples. Further details of this extended NRE framework are provided in Appendix~\ref{subsec:extended_NRE}.

\section{Implementation, verification and benchmarking}
\begin{figure}[t]
  \centering
  \includegraphics[width=.52\textwidth]{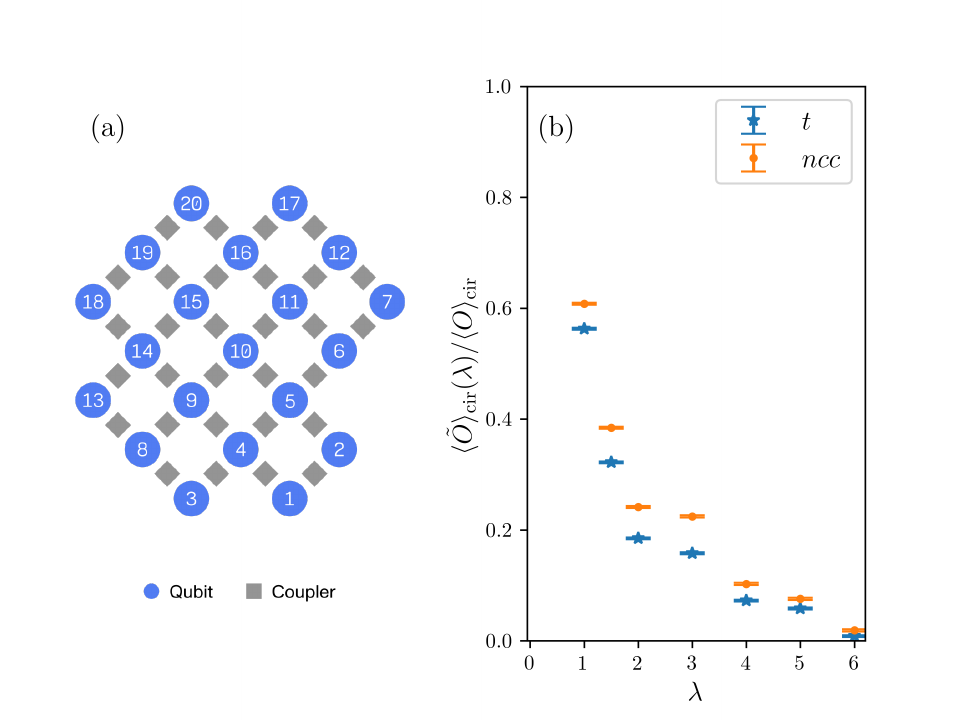}
\caption{(a) Schematic of the IQM Garnet, a 20-qubit superconducting quantum processor with square-grid connectivity. 
(b) Measured ground-state energy of the transverse-field Ising model as a function of the noise scale factor $\lambda$, normalized to the noiseless expectation value. Results are shown for both the target circuit (\textit{t}) and the noise-canceling circuit (\textit{ncc}).}
  \label{fig:TFIM_noisy_expts}
\end{figure}
\begin{figure*}
    \centering
\includegraphics[width=1.\linewidth]{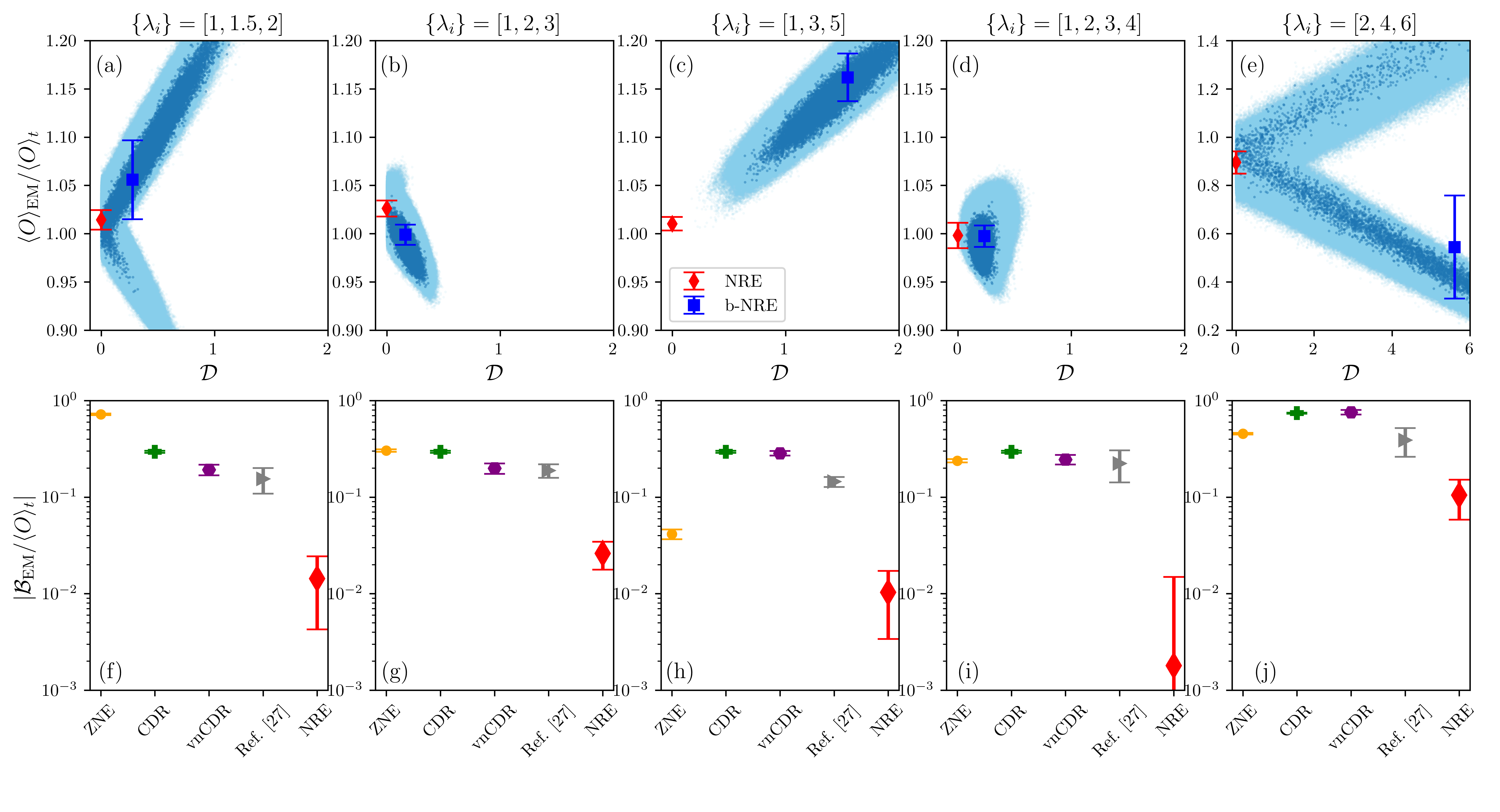}
\caption{(a–e) Noise-Robust Estimation of the TFIM ground-state energy for different sets of noise scale factors. 
Dark blue shaded regions show $2\times10^{4}$ baseline estimations obtained by resampling the first bootstrap set ($s{=}1$) of expectation values using the extended NRE procedure (see Appendix ~\ref{subsec:extended_NRE}). 
Light blue shaded regions aggregate the corresponding baseline estimations across all 500 bootstrap sets. 
Red points indicate the mean and standard deviation of the final NRE estimations, while blue points show the mean and standard deviation of the baseline estimations; in both cases error bars are derived from 500 bootstrap samples. 
Panel (c) corresponds to the dataset shown in Fig.~\ref{fig:NRE_Schematics}(c). 
(f–j) Relative bias magnitude $|\mathcal{B}_\mathrm{EM}/\langle O\rangle_t|$ after applying different error-mitigation strategies, using the same noise scale factors as in (a–e). 
All methods are compared under a fixed budget of $2.4\times10^{5}$ measurement shots. 
Because CDR does not require noise amplification (unlike vnCDR), its predictions remain constant across panels (f–i). 
In panel (j), where $\lambda_{1}=2$, the CDR ansatz is trained using noisy measurements of the training circuits at $\lambda_1=2$.}
    \label{fig:NRE_on_Garnet}
\end{figure*}

We implemented the NRE framework and evaluated its performance on the \textit{IQM Garnet} processor~\cite{Garnet}, a 20-qubit superconducting quantum processing unit with square-grid connectivity, as illustrated in Fig.~\ref{fig:TFIM_noisy_expts}(a). In all experiments presented below, noise-canceling circuits are constructed from the corresponding target circuits by replacing each non-Clifford gate with the Clifford gate that minimizes the Frobenius norm of the unitary difference between the two gates. Further details on the construction of noise-canceling circuits are provided in Appendix~\ref{sec:ap_ncc}.

\subsection{Transverse-field Ising model}
\label{sec:TFIM_implementation}

As a first case study, we consider error mitigation for estimating the ground-state energy of the transverse-field Ising model (TFIM),
\begin{align}
\label{eq:tfi_hamiltonian}
H_{\text{TFIM}} = -g\sum_{j}\sigma_X^j - \sum_{(i,j)}\sigma_Z^i\sigma_Z^{j}.
\end{align}
We focus on the paramagnetic regime with transverse field strength $g=2$ on the square-grid connectivity shown in Fig.~\ref{fig:TFIM_noisy_expts}(a). The ground-state energy is estimated using a QAOA ansatz with $p=4$ layers. The variational parameters are first optimized in noiseless simulation, and the resulting optimized circuit is then executed on the quantum processor as the target circuit for the mitigation experiments.

The target circuits executed on the IQM backend use $20$ qubits, contain $240$ entangling CZ gates, and have depths of $76$ and $77$ for measurements in the $Z$ and $X$ bases, respectively.

To access different effective noise strengths, we employ global unitary folding~\cite{Schultz2022}, which allows controlled variation of the noise scale factor $\lambda$ and its spacing $h$. Figure~\ref{fig:TFIM_noisy_expts}(b) shows the noisy expectation values obtained from the target and noise-canceling circuits at different noise scale factors, normalized by their respective noiseless values. Already at the smallest noise scale factor, $\lambda_1=1$, the measured expectation value of the target circuit is reduced by more than $40\%$ relative to the ideal value.

Figure~\ref{fig:NRE_on_Garnet} summarizes the NRE results for different choices of noise-scale spacing, number of noise scales, and circuit depth. For each experimental setting, we used $2\times10^{4}$ measurement shots per quantum circuit execution on the Garnet quantum processor. Consequently, NRE with three noise scale factors required a total shot budget of $2.4\times10^{5}$. In all panels of Fig.~\ref{fig:NRE_on_Garnet}, when needed, we rescaled the $1\sigma$ error bars of the other error-mitigation methods so that they correspond to the same total shot budget.

To perform NRE, we generated $500$ bootstrap resamples from the resulting count data, from which bootstrapped expectation values were computed. Each bootstrapped expectation value was then further resampled $2\times10^{4}$ times according to the extended NRE procedure described in Appendix~\ref{subsec:extended_NRE}.

Figures~\ref{fig:NRE_on_Garnet}(a)–(c) correspond to noise scale factors of the form $\{\lambda_i\}=[1,1+h,1+2h]$ with $h=0.5$, $1$, and $2$, respectively. In Fig.~\ref{fig:NRE_on_Garnet}(d), the number of noise scales is increased to $M=4$ while keeping $h=1$. Figure~\ref{fig:NRE_on_Garnet}(e) uses $\{\lambda_i\}=[2,4,6]$, which has the same relative spacing as panel (b) but approximately doubles the circuit depth. This places the experiment in a substantially noisier regime. In particular, the circuits at $\lambda=2$ for the $Z$- and $X$-basis measurements each contain $480$ entangling CZ gates, and the corresponding target expectation value $\langle\tilde{O}\rangle_t(\lambda=2)$ is suppressed by approximately $80\%$ relative to its noiseless value. In all panels, the shaded regions indicate the distributions of baseline estimators obtained from bootstrapping and resampling.

Across all tested settings, baseline estimators with smaller normalized dispersion $\mathcal{D}$ consistently exhibit smaller residual bias magnitude $|\mathcal{B}_{\mathrm{b\text{-}NRE}}|$ and therefore lie closer to the ideal expectation value. This behavior is consistent with the analytical framework developed in Sec.~\ref{sec:NRE_workflow} and Appendix~\ref{sec:ap_residual_bias}, where the residual bias and the variation of the auxiliary estimator across noise scales are governed by the same hierarchy of derivatives. The normalized dispersion therefore provides a practically measurable proxy for the magnitude of the residual bias of the baseline estimator.

In several cases, most clearly in Figs.~\ref{fig:NRE_on_Garnet}(a) and (e), the baseline estimators split into two branches with positive and negative residual bias. This bifurcation shows that $\mathcal{D}$ captures the \emph{magnitude} of the residual bias but not its sign: estimators with similar dispersion can have opposite bias signs while retaining comparable bias magnitude. This is again consistent with the analysis of Appendix~\ref{sec:ap_residual_bias}, where distinct combinations of higher-order derivative contributions can reverse the sign of the bias without substantially changing its overall scale.

To obtain the final NRE estimate, we perform a weighted linear regression over the set of baseline estimators, assigning each $\langle O \rangle_{\mathrm{b\text{-}NRE}}^{\,s}$ a weight proportional to $1/\mathcal{D}_s$. This emphasizes low-dispersion baseline estimators, which are expected to have smaller residual bias. As discussed in Appendix~\ref{sec:ap_regression_strategies}, the final NRE estimate is largely robust to the precise choice of weighting scheme.

In several settings, this second post-processing step substantially reduces the statistical uncertainty relative to the baseline estimator. The effect is most pronounced when the baseline estimators span a wide range of dispersion values: the regression suppresses high-dispersion realizations, which are both noisier and more strongly biased, while giving greater weight to low-dispersion realizations. In this sense, the extrapolation toward the $\mathcal{D}\to 0$ limit also acts as a structured variance-reduction procedure. By contrast, when the baseline estimators are already concentrated at small $\mathcal{D}$, the second post-processing step yields only a modest improvement, since both the residual bias and the statistical spread are already limited.

In Figs.~\ref{fig:NRE_on_Garnet}(f–j), we compare the residual bias of NRE with several noise-agnostic error-mitigation methods. We report the relative bias magnitude $|\mathcal{B}_\mathrm{EM}/\langle O\rangle_t|$ and benchmark against ZNE~\cite{IBM_ExpZNE2019}, the method of \ocite{Urbanek_2021}, CDR~\cite{Czarnik_2021}, and vnCDR~\cite{Lowe2021}. For ZNE, we use a single-exponential fit. The implementation of the method of \ocite{Urbanek_2021} uses the same noise-canceling circuit as the estimation circuit introduced in that work. For CDR and vnCDR, we generate $40$ near-Clifford training circuits by replacing approximately $90\%$ of the non-Clifford gates in the target circuit with Clifford gates, following Ref.~\cite{Czarnik_2021}.

As shown in Figs.~\ref{fig:NRE_on_Garnet}(f–h), ZNE is highly sensitive to the choice of noise scale factors. Because the optimal set of scale factors is generally not known \emph{a priori}, this sensitivity limits its practical robustness. Among the tested cases, ZNE performs best in panel (h), where the noise scale factors are odd integers; a plausible explanation is that gate-based noise amplification is implemented more consistently for these values. In contrast, NRE yields comparatively stable estimates across all tested noise-scale configurations.

Finally, Fig.~\ref{fig:NRE_on_Garnet}(j) corresponds to a substantially noisier regime in which even the smallest noise scale factor causes strong suppression of the target expectation value. In this setting, all other error-mitigation methods considered here fail to recover an accurate estimate, whereas NRE remains the most reliable and recovers an estimate within approximately $10\%$ of the ideal expectation value.

\subsection{\(\operatorname{H}_4\) molecule}
\label{sec:H4}
\begin{figure}
    \centering
    \includegraphics[width=1.1\linewidth]{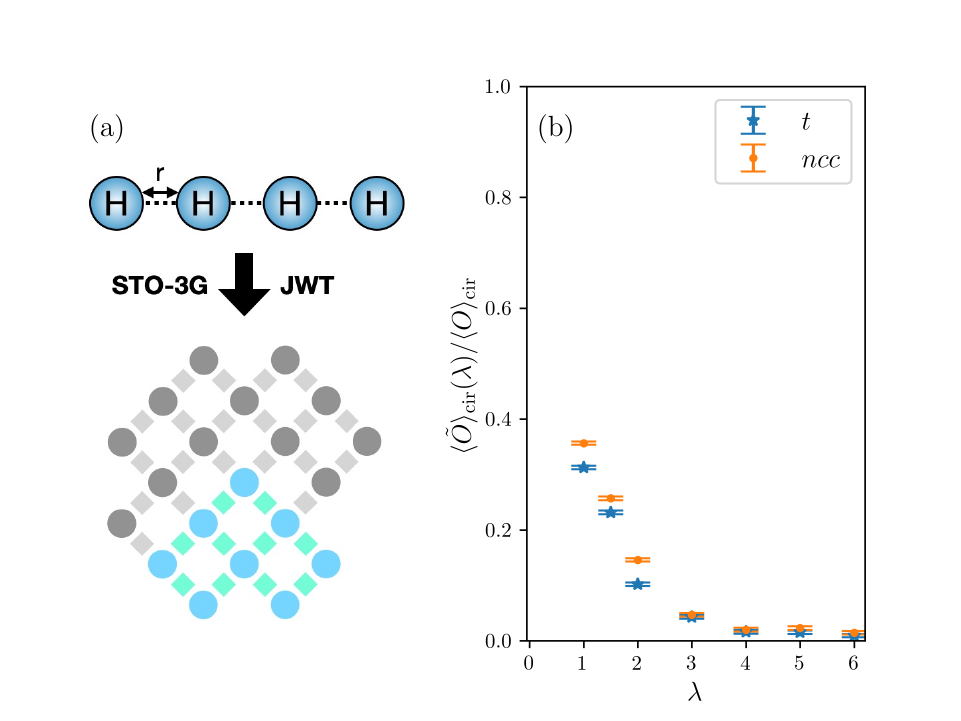}
\caption{(a) Geometry of the $\mathrm{H}_4$ molecule in the STO-3G basis set. The target circuits are executed on the subset of 8 qubits highlighted in blue on the processor. 
(b) Measured expectation values of the $\mathrm{H}_4$ energy observable as a function of the noise scale factor $\lambda_i$, normalized to the corresponding noiseless values, 
$\langle \tilde{O} \rangle_{\mathrm{cir}}(\lambda_i) / \langle O \rangle_{\mathrm{cir}}$, 
for both the target circuit (\textit{t}) and the corresponding noise-canceling circuit (\textit{ncc}).}
\label{fig:NRE_fidelity_H4}
\end{figure}
\begin{figure*}
    \centering
    \includegraphics[width=1\linewidth]{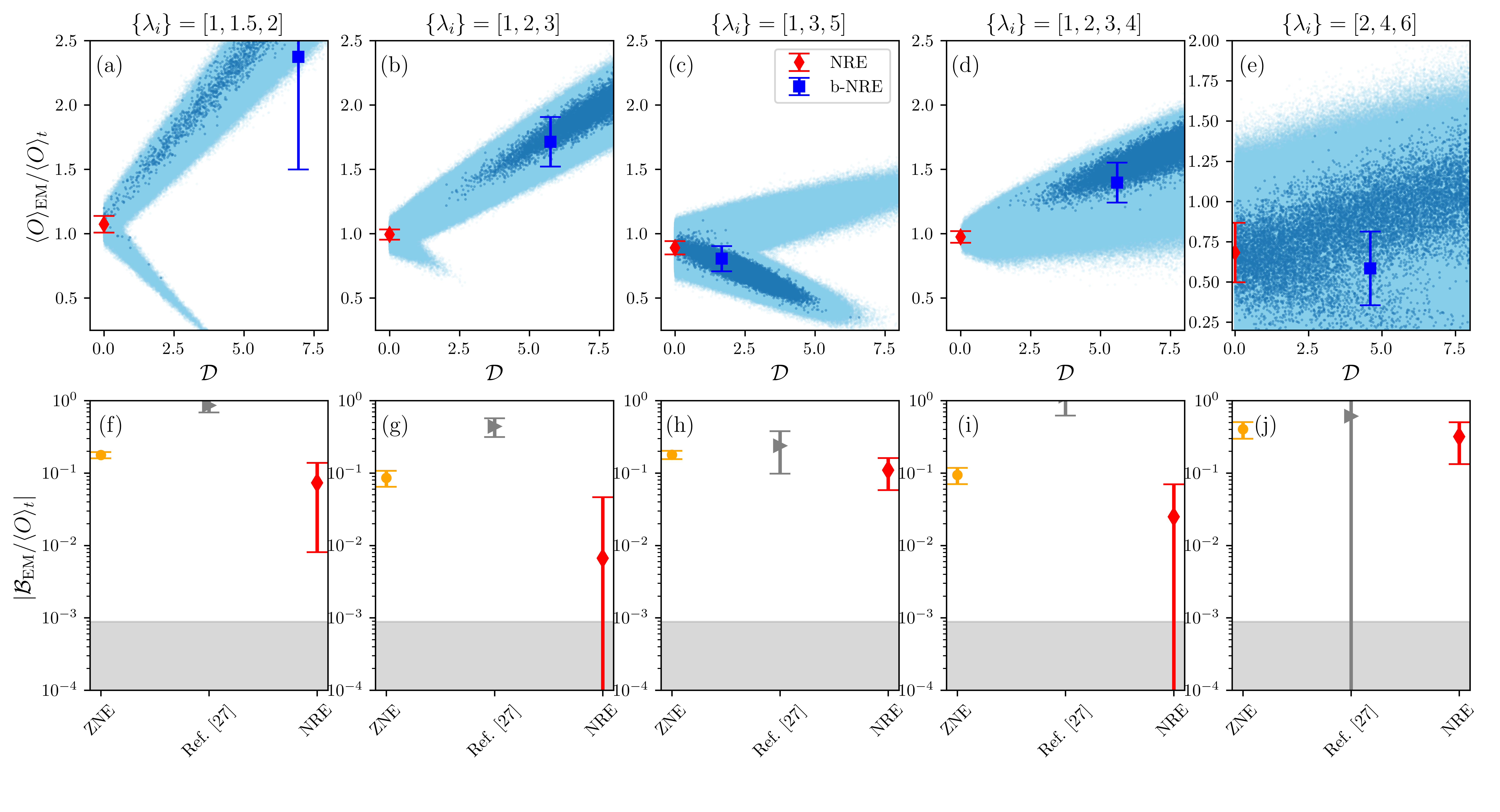}
\caption{(a–e) Noise-Robust Estimation of the ground-state energy of the H$_4$ molecule for different choices of noise scale factors. 
As in Fig.~\ref{fig:NRE_on_Garnet}, dark blue shaded regions show $2\times10^{4}$ baseline estimations obtained by resampling the first bootstrap set ($s{=}1$), while light blue shaded regions aggregate the corresponding baseline estimations across all 500 bootstrap sets. 
(f–j) Relative bias magnitude after applying different error-mitigation strategies, including ZNE, the method of Ref.~\cite{Urbanek_2021}, and NRE. 
The shaded area marks the threshold for chemical precision.}
\label{fig:NRE_Benchmarks_H4}
\end{figure*}
While the TFIM Hamiltonian considered above has a relatively sparse and structured form, many practical applications involve Hamiltonians with significantly denser interaction structure. To assess the performance of NRE in such settings, we next consider electronic structure Hamiltonians arising in quantum chemistry.

The electronic structure Hamiltonian takes the form

\begin{equation}
\mathcal{H}_{\text{ES}} = \sum_{pq}h^{pq} c^\dagger_pc_q + \sum_{pqrs}h^{pqrs} c^\dagger_p c^\dagger_q c_r c_s ,
\end{equation}
where $p,q,r,s$ label basis functions, $c^\dagger$ and $c$ are fermionic creation and annihilation operators, and $h^{pq}$ and $h^{pqrs}$ denote one- and two-electron integrals. In contrast to the TFIM Hamiltonian, where the number of interaction terms scales as $O(n^2)$ with the number of qubits $n$, the electronic structure Hamiltonian contains $O(n^4)$ interaction terms and is therefore substantially denser.

As a benchmark system, we consider the hydrogen chain $\operatorname{H}_4$ with inter-atomic distance $1.0\,\angstrom$. Using the minimal STO-3G basis set, the fermionic system is mapped onto $8$ qubits via the Jordan–Wigner transformation.

The ground-state energy is approximated using a classically pre-trained variational quantum eigensolver (VQE) circuit \cite{tilly2022variational} employing the unitary paired coupled cluster singles and doubles (UpCCSD) ansatz \cite{lee2018generalized}. In noiseless simulation this ansatz achieves an energy of $-2.1351$ Ha, compared with the exact FCI value of $-2.1663$ Ha.

The $185$ Hamiltonian terms are grouped into $8$ commuting operator sets \cite{yen2020measuring,verteletskyi2020measurement}, allowing simultaneous measurement after a global basis rotation. In these rotated bases, the Hamiltonian operators may still be highly nonlocal, with Pauli weights up to $6$. The required rotation unitaries increase the circuit depth by $O(N)$ layers and introduce $O(N^2)$ two-qubit gates. As a result, the transpiled circuits used for energy estimation have depths between $132$ and $264$, corresponding to between $89$ and $169$ CZ gates.

Compared with the TFIM example, this configuration results in deeper circuits and therefore more rapid decay of the measured expectation values as the noise scale factor increases, as illustrated in Fig.~\ref{fig:NRE_fidelity_H4}. We observe that $\langle\tilde{O}\rangle_t(\lambda=1)$ is reduced by approximately $70\%$ relative to its noiseless value, whereas for $\lambda>3$ the measured expectation values for both the target and noise-canceling circuits approach approximately $2\%$ of their noiseless values, indicating that most information about the quantum state has been lost.

Despite this strongly noisy regime, the NRE framework is still able to restore the estimated energy close to the ideal value, as shown in Fig.~\ref{fig:NRE_Benchmarks_H4}. The corresponding noise-canceling circuits are constructed using the same procedure as in the TFIM experiments.

As in the previous example, we considered five different sets of equally spaced noise-scale configurations. For each experimental setting, we used $2\times10^{4}$ measurement shots per quantum circuit execution on the processor. Consequently, NRE with three noise scale factors required a total shot budget of $9.6\times10^{5}$. In all panels of Fig.~\ref{fig:NRE_Benchmarks_H4}, when needed, we rescaled the $1\sigma$ error bars so that they correspond to the same total shot budget. For each configuration, we generated $500$ bootstrap resamples from the experimental data, followed by $2\times10^{4}$ resamples per bootstrap set in the extended NRE procedure.

For several noise-scale configurations (panels a, b, and d), the estimated energy lies within the statistical uncertainty of the ideal VQE result. The most accurate estimate occurs for the noise-scale set $\{\lambda_i\}=[1,2,3]$, where the relative error is approximately $0.66\%$. The corresponding uncertainty is approximately $3.9\%$ of the ratio between the extrapolated and exact VQE energies, which lies close to the threshold of chemical precision ($0.00159$ Ha), indicated by the shaded region in Fig.~\ref{fig:NRE_Benchmarks_H4}.

Consistent with the TFIM experiment, we again observe a strong correlation between baseline estimation accuracy and normalized dispersion, indicating that the $|\mathcal{B}_\mathrm{b\text{-}NRE}|$–$\mathcal{D}$ relationship also persists in this more complex Hamiltonian setting.

For this molecular benchmark we restrict the comparison to ZNE and the method of \ocite{Urbanek_2021}. Because estimating the $\operatorname{H}_4$ energy requires executing $8$ measurement circuits, applying CDR with the same number of training circuits used in Sec.~\ref{sec:TFIM_implementation} would require $320$ training circuit executions, while vnCDR would require even more depending on the number of noise scales used.

We note that the relatively lower accuracy of NRE for the setting shown in panel (c), compared with panels (a) and (b)—which also use three noise scale factors—is not accidental. As shown in Fig.~\ref{fig:NRE_fidelity_H4}(b), at $\lambda_i = 5$ the target expectation value is almost completely suppressed by noise and approaches zero. In this regime the measured observable carries little information about the ideal value. In contrast, for panels (a) and (b), which use smaller noise-scale spacing, the largest noise scale factor still allows $\tot(\lambda_3)$ to retain partial information about the ideal expectation value.

This observation highlights an important practical consideration: for strongly noisy circuits it is advantageous to use smaller noise-scale factors when implementing gate-based noise amplification. At the same time, the NRE framework remains relatively robust to imperfections in the realized noise amplification.

Figure~\ref{fig:NRE_Benchmarks_H4} shows that, across the tested settings, NRE consistently yields estimates closer to the ideal value than the other mitigation strategies considered here. Panels (e) and (j) correspond to an extremely noisy regime where even at the smallest noise scale factor ($\lambda_1=2$) the target expectation value is suppressed by approximately $90\%$. Despite this severe noise degradation, NRE still recovers an estimate closer to the ideal value, with a relative accuracy of approximately $70\%$.

\begin{figure*}
    \centering\includegraphics[width=1\linewidth]{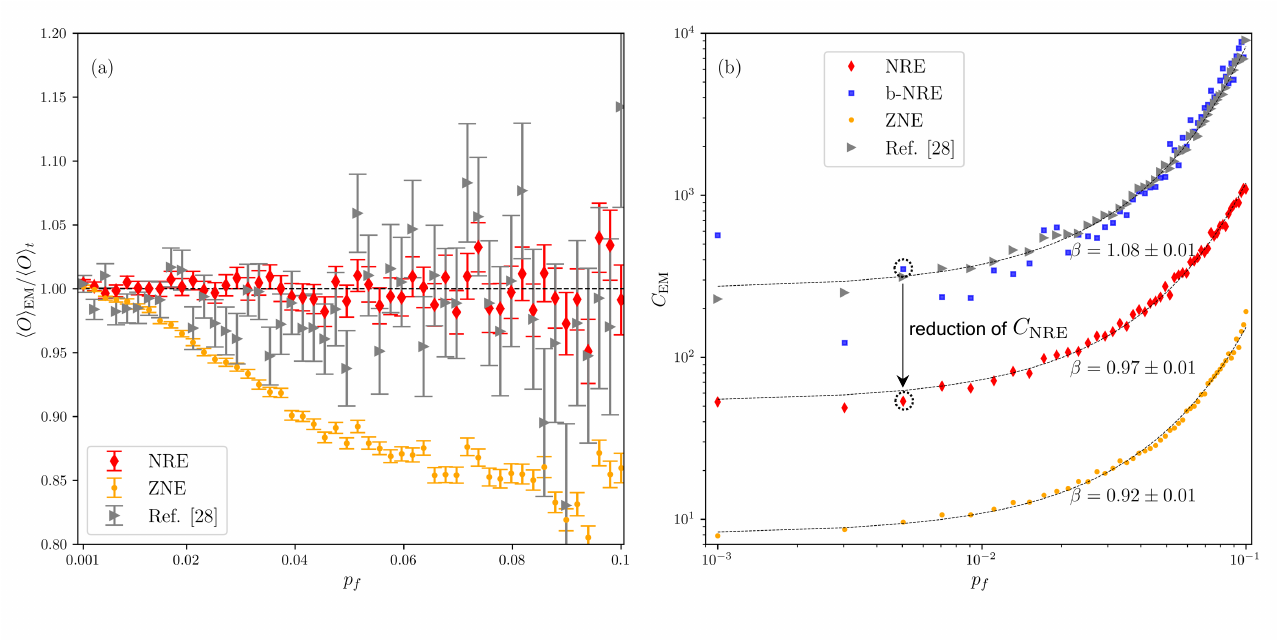}
\caption{(a) Simulated error-mitigated expectation values, normalized to the noise-free value, as a function of the two-qubit gate error rate $f$. Error bars are computed using a fixed total measurement budget of $6\times10^5$ shots for all mitigation methods. Bootstrapping is performed $2000$ times for each simulated dataset, and each bootstrapped expectation value is further resampled $10^4$ times.
(b) Sampling overhead $C_\mathrm{EM}$ as a function of the two-qubit gate error rate $f$. The overhead is computed according to Eq.~(\ref{eq:C_EM}), using the variance of $\langle \tilde{O} \rangle_t(\lambda_1)$ obtained with the same total shot budget as the mitigated estimator $\langle O \rangle_\mathrm{EM}$. The circuit contains $N_\mathrm{TQG}=32$ two-qubit gates. Dashed lines show exponential fits of the form given in Eq.~(\ref{eq:C_EM_app}) for each mitigation method.}
\label{fig:Sampling_overhead}
\end{figure*}

\section{Sampling overhead analysis}
\label{subsec:SO}

In this section, we analyze the sampling overhead associated with NRE through numerical simulations. It is well known that error-mitigation methods generally introduce additional sampling overhead, which manifests as an increase in the variance of the mitigated estimator relative to the noisy estimator at a given circuit error rate $\epsilon_0$ \cite{Endo_2021, Cai_2023}. The sampling overhead can be quantified as

\begin{align}
\label{eq:C_EM}
C_\mathrm{EM} =
\frac{\mathrm{Var}[\langle O \rangle_\mathrm{EM}]}
{\mathrm{Var}[\langle \tilde{O} \rangle_t(\lambda_1)]}.
\end{align}

Assuming, for simplicity, a Pauli noise model in which the error probability per noisy gate operation is $f$ and is uniformly distributed across the circuit, the total circuit error rate can be approximated as $\epsilon_0 = Nf$, where $N$ denotes the total number of gate operations in the circuit. Under this assumption, the sampling overhead can often be approximated by the exponential form

\begin{align}
\label{eq:C_EM_app}
C_\mathrm{EM} \approx \alpha e^{\beta Nf},
\end{align}

where the parameters $\alpha$ and $\beta$ depend on the specific error-mitigation method and its implementation. Exponential scaling of the sampling overhead with circuit size and error rate has been extensively studied in the literature \cite{Cai_2023, Takagi_2023, Tsubouchi_2023, Quek_2024, Filippov_2024}.

To investigate the sampling overhead of NRE, we simulate the same TFIM problem introduced in Sec.~\ref{sec:TFIM_implementation}. For computational tractability, we consider a reduced $5$-qubit system with the connectivity map of the IQM Spark device \cite{IQM_Spark}, consisting of four outer qubits each coupled to a central qubit.

We employ a simplified noise model in which all two-qubit gates experience local depolarizing noise with rate $f$, while single-qubit gates are assumed to be ideal. The two-qubit gate error rate $f$ is varied within the range $[10^{-3},\,10^{-1}]$. This simplified noise model is chosen to isolate the statistical scaling behavior of the mitigation methods and should therefore be interpreted as a controlled study of sampling overhead rather than a detailed model of hardware noise. While realistic quantum devices exhibit more complex noise processes, the exponential dependence of sampling overhead on circuit error rate is expected to persist for a broad class of noise models \cite{Cai_2023, Takagi_2023}.

Figure~\ref{fig:Sampling_overhead}(a) shows the mitigated expectation values as a function of the two-qubit gate error rate $f$ for NRE, ZNE, and the depolarizing-noise mitigation method proposed in Ref.~\cite{Urbanek_2021}. As in the previous sections, the final NRE estimate is obtained through the weighted regression step described in Sec.~\ref{sec:TFIM_implementation}. The total number of measurement shots is kept fixed across all three methods, and three noise scale factors $\{\lambda_i\}=[1,2,3]$ are used. To reduce statistical fluctuations arising from finite sampling in the numerical simulations, we increase the number of bootstrap resamplings to $2000$ when applying NRE in this section.

As shown in Fig.~\ref{fig:Sampling_overhead}(a), the accuracy of ZNE deteriorates at higher two-qubit error rates. In contrast, NRE maintains accurate estimates across a wider range of error rates, with its uncertainty primarily limited by statistical fluctuations arising from the finite number of measurement shots. It is worth noting that the local depolarizing noise model considered here represents a regime in which the method of Ref.~\cite{Urbanek_2021} is expected to perform favorably. Nevertheless, NRE produces estimates with substantially smaller statistical uncertainties over a wide range of error rates.

Figure~\ref{fig:Sampling_overhead}(b) shows the numerically computed sampling overhead $C_\mathrm{EM}$, defined in Eq.~(\ref{eq:C_EM}), for each mitigation method. We also include the sampling overhead of the baseline estimator used in the first step of NRE, which shows behavior similar to the method of Ref.~\cite{Urbanek_2021}. The dashed curves correspond to exponential fits of the form given in Eq.~(\ref{eq:C_EM_app}), from which the parameter $\beta$ is extracted for each method.

Consistent with the observations from the experimental results presented earlier, the second post-processing step of NRE—which exploits the $|\mathcal{B}_{\mathrm{b\text{-}NRE}}|$–$\mathcal{D}$ correlation—significantly reduces the sampling overhead relative to the baseline estimator. In the example shown in Fig.~\ref{fig:Sampling_overhead}(b), the average sampling overhead of NRE is reduced by a factor of approximately $6.2$ compared with the baseline estimation.

An additional observation is that the regression step of NRE not only reduces the prefactor of the sampling overhead but also slightly decreases the exponential scaling coefficient $\beta$ relative to the baseline estimator. This indicates that the dispersion-based filtering mechanism preferentially selects baseline realizations with smaller variance amplification, thereby improving the effective scaling of the estimator variance with circuit error rate. Although the improvement in $\beta$ is modest in the present example, even small reductions in the exponential scaling coefficient can translate into substantial savings in sampling cost for larger circuits.

For the specific problem studied here, the sampling overhead of NRE is on average about a factor of $7$ larger than that of ZNE. However, because NRE requires executing both the target circuit and the noise-canceling circuit, the same statistical precision can be achieved by increasing the number of shots per circuit by only a modest factor of approximately $3.5$. These results illustrate the practical trade-off between variance amplification and bias suppression in error mitigation: while ZNE exhibits smaller variance amplification, the residual bias at larger error rates limits its accuracy, whereas NRE maintains reliable estimates with only a moderate increase in sampling overhead.

Finally, we emphasize that the fitted parameters $\alpha$ and $\beta$ depend on several problem-specific factors, including the observable being measured and the number and spacing of noise scale factors. For comparison, in probabilistic error cancellation (PEC) these parameters typically take the values of $\alpha = 1$ and $\beta = 4$ \cite{Cai_2023}.

\section{\label{sec:conclusion}Discussion and outlook}

In this work we introduced Noise-Robust Estimation (NRE), a noise-agnostic
error-mitigation framework for expectation-value estimation on noisy quantum
hardware. The method combines a target circuit with a structurally matched
noise-canceling companion circuit and a two-stage post-processing strategy to
construct a baseline estimator with reduced noise sensitivity and then further
suppress its residual bias through extrapolation to the limit of vanishing
normalized dispersion. Our results show that this approach can achieve strong
bias suppression while maintaining a moderate sampling overhead.

A central observation underlying NRE is that the magnitude of the residual bias
of the baseline estimator is correlated with the normalized dispersion of the
auxiliary estimator across amplified noise realizations. Within the NRE
formalism, this dispersion is not merely a heuristic indicator, but a
measurable diagnostic of the residual-bias magnitude of the baseline
estimator. As discussed in Appendix~\ref{sec:ap_residual_bias}, both
quantities are governed by the same underlying derivative structure of the
auxiliary estimator with respect to the noise scale. This provides a natural
interpretation of why extrapolation to the $\mathcal{D}\to 0$ limit improves
the final estimate.

Experimentally, we found that NRE consistently outperforms several established
noise-agnostic mitigation methods across different noise-amplification
settings and for both low-weight observables, such as the TFIM ground-state
energy, and high-weight observables, such as the ground-state energy of
$\mathrm{H}_4$. In the latter case, NRE yields estimates close to chemical
accuracy even when the measured energy is strongly suppressed relative to its
noiseless value. In addition, the second post-processing step not only reduces
the residual bias of the baseline estimator but can also reduce its
statistical uncertainty by suppressing the influence of baseline realizations
with large normalized dispersion.

Another notable feature of the protocol is its robustness with respect to
algorithmic hyperparameters, such as the spacing of the noise levels and the
regression strategy used in the final post-processing step. In contrast to ZNE,
whose performance can depend strongly on the extrapolation model and the
choice of noise scale factors, we find that NRE exhibits only weak sensitivity
to both the noise-level spacing and the precise form of the regression used to
extrapolate to the $\mathcal{D}\!\to\!0$ limit. As shown in
Appendix~\ref{sec:ap_regression_strategies}, different weighting schemes lead
to very similar final estimates over a broad range of parameters. At the same
time, the present experiments also indicate the limits of the current
protocol: in extremely noisy regimes, such as those shown in
Figs.~\ref{fig:NRE_on_Garnet}(e) and \ref{fig:NRE_Benchmarks_H4}(e), the
accuracy of the final estimation is reduced, although NRE still remains more
reliable than the alternative methods considered in this work.

We also note that our experiments presented in this paper did \textit{not} involve additional error reduction methods such as readout error mitigation \cite{Nachman_2020, Nation_2021, Yang_2022}, randomized compiling \cite{Akel_2021}, or dynamical decoupling \cite{Ezzell_2023}. This demonstrates that, in the presented experiments, NRE functioned as a stand-alone error mitigation method capable of providing accurate estimations.

This indicates that NRE can already provide accurate estimations as a stand-alone mitigation protocol, even in the presence of readout errors and coherent error sources. Because these error sources affect both the target and noise-canceling circuits, part of their effect can be absorbed into the auxiliary-estimator construction and subsequent post-processing. Nevertheless, combining NRE with dedicated error-suppression techniques is expected to further improve performance, especially for deeper circuits. Combining NRE with other error suppression methods represents a promising avenue for improving its effectiveness.

Several directions for further development follow naturally from these results.
On the methodological side, an important area for refinement is the second
post-processing layer. While the present results are robust across different
weighting schemes (see Appendix~\ref{sec:ap_regression_strategies}), the
extrapolation to the zero-dispersion limit could be improved through more
flexible regression strategies. A particularly promising direction is the use
of nonparametric, uncertainty-aware models, such as Gaussian Process
Regression (GPR), to infer the extrapolated value together with a principled
uncertainty estimate. Such approaches may be especially useful in high-noise
regimes, where the relation between the baseline estimator and the dispersion
can become less linear and the bootstrap samples more unevenly distributed.
More generally, machine-learning models that incorporate richer features than
$\mathcal{D}$ alone---for example, features derived from the shape of
$\mathcal{A}(\lambda_i)$ across noise scales, local curvature, asymmetry, or
additional diagnostics quantifying mismatch between the target and
noise-canceling circuits---may help resolve branch structure in the baseline
cloud and further improve the stability of the final extrapolation. Machine learning has already been explored in previous studies for error mitigation \cite{Bennewitz_2022, Liao_2024}, and similar approaches could improve the adaptability and effectiveness of NRE.

A second natural direction concerns the construction of the noise-canceling
circuit itself. In the present work, the \textit{ncc} is obtained by a local
``closest Clifford'' replacement strategy based on the Frobenius norm.
Although this approach already yields accurate results, future work could treat
\textit{ncc} construction as a global optimization problem. In particular, one
could seek families of candidate \textit{ncc}s for a single target circuit
that improve the stability of the $\mathcal{D}\to 0$ extrapolation, enlarge
the useful spread in dispersion values, or reduce systematic mismatch between
the noise responses of the target and auxiliary circuits. Here, machine
learning may again play a useful role: one can envision data-driven or
reinforcement-learning-based approaches that search over candidate Clifford
replacements or circuit-level transformations and optimize objective functions
built from dispersion spread, extrapolation stability, and agreement across
multiple candidate \textit{ncc}s. Such an ensemble-based approach could also
provide an internal consistency test: agreement among extrapolated results
obtained from multiple \textit{ncc} constructions would strengthen confidence
in the robustness of the protocol.

Finally, as quantum processors progress toward fault tolerance, hybrid
scenarios combining error mitigation with partial quantum error correction are
likely to become increasingly relevant~\cite{Piveteau_2021,
Suzuki_2022}. Such combinations have already been explored for other
mitigation strategies, including quasi-probabilistic mitigation and
ZNE~\cite{Lostaglio_2021, Wahl_2023, Zhang_2025, zhou_2025, Jeon_2026}. Recent work has further
shown that logical-level error mitigation can be performed efficiently by
exploiting soft information naturally produced by QEC decoders, enabling
in situ characterization and mitigation of logical operations without
separate characterization experiments or additional quantum runtime beyond
standard QEC workflows~\cite{zhou_2025}. In this context, it is natural
to ask whether NRE can also be extended to the logical level, where
noise-canceling circuits would need to be designed for encoded operations and
logical observables, and where decoder-derived logical information could
potentially be combined with the NRE bias-diagnostic framework. Such a hybrid
NRE--QEC strategy would be particularly attractive because the central idea of
NRE---using an observable statistical diagnostic to track residual-bias
magnitude without explicit microscopic noise modeling---is conceptually
compatible with logical-level soft-information approaches. Exploring these
directions could extend the applicability of NRE far beyond the NISQ era and
constitutes an important direction for future research.

In conclusion, NRE provides a versatile approach to error
mitigation that is robust to imperfections in noise amplification and does not rely on explicit noise modeling. By combining structured auxiliary circuits
with bias diagnostics derived directly from measurement data, the framework
opens a path toward more reliable quantum computations on near-term devices
and offers promising directions for extension to more advanced quantum
computing architectures.

\section*{Data availability}
All experimental and simulation data supporting the findings of this paper are available from the corresponding author upon reasonable request.
\section*{Acknowledgements}
We thank all employees at IQM Quantum Computers for their insightful discussions. We especially thank Stefan Pogorzalek, St\'ephanie Cheylan, Martin Leib, Francisco Revson F. Pereira, Miha Papi\v c, Pedro Figueroa-Romero, Manish Thapa, Hao Hsu,  and 	
Raphael A. Brieger. Additionally, we acknowledge the support from the German Federal Ministry of Education and Research (BMBF) under QSolid (grant No. 13N16161).
\section*{Author Contributions}
A.~H. conceived the project, developed the error-mitigation framework,
performed the numerical simulations and experimental verification, and
wrote the manuscript. F.~S. proposed the quantum chemistry simulation of
the $\mathrm{H}_4$ molecule, wrote the hydrogen-chain section of the
manuscript, and contributed to the conceptual development of the method.
A.~C. contributed to the development of the computer code for
bootstrapping, assisted in extending the framework to include uncertainty
estimation, and contributed to the early conceptual development of the
project. T.~L. and A.~A. contributed to the early conceptual development
of the method. J.~D.~G., E.~G.~R., and V.~P~C. assisted in developing the computer codes. I.~d.~V. facilitated the project and contributed to
the early conceptual development of the work. All authors reviewed,
revised, and approved the final manuscript.
\section*{Competing Interests}
The authors declare the following competing interests: The error mitigation framework presented in this work is part of patent application FI20246514, with A. H. listed as the inventor.

\appendix
\section{Analytical Structure of the Residual Bias}
\label{sec:ap_residual_bias}

In this appendix we analyze the analytical structure of the residual bias in the baseline NRE estimator and identify the contributions arising from truncation error, discretization error, and imperfect noise amplification. We further show how these contributions are naturally connected to the normalized dispersion introduced in the main text.

For quantum circuits containing non-Clifford gates and subject to Pauli noise, the expectation value of a Pauli observable generally takes the form of a sum of exponential functions of the noise scale parameter $\lambda$. For Clifford circuits this dependence reduces to a single exponential decay. Under the assumption that $\langle \tilde{O} \rangle_{ncc}(\lambda)$ remains nonzero and of fixed sign over the range of noise scales considered, the auxiliary quantity $\mathcal{A}(\lambda)$ defined in Eq.~(\ref{eq:aux_O}) is therefore a smooth function of $\lambda$.

Let $\mathcal{A}^{(j)}_{\lambda_1}$ denote the $j$th derivative $\partial_\lambda^{\,j}\mathcal{A}(\lambda)$ evaluated at $\lambda_1=1$. Although $\mathcal{A}$ and its derivatives depend on the control parameter $n$ and the noise-canceling circuit, these dependencies are omitted for clarity. Expanding $\mathcal{A}(\lambda)$ around $\lambda_1$ and evaluating at $\lambda=0$ yields
\begin{align}
\label{eq:Taylor_summ_te}
\langle O\rangle_t
=
\mathcal{A}(\lambda_1)
+
\sum_{j=1}^{M-1}
\frac{(-\lambda_1)^j}{j!}\,
\mathcal{A}^{(j)}_{\lambda_1}
+
\mathcal{T}(M),
\end{align}
where
\begin{align}
\mathcal{T}(M)
=
\sum_{k=M}^{\infty}
\frac{(-\lambda_1)^k}{k!}
\mathcal{A}^{(k)}_{\lambda_1}
\end{align}
represents the truncation error arising from higher-order derivatives that cannot be estimated from the available data.

Since only $M$ noise-scale data points are available, derivatives can only be numerically estimated up to order $M-1$, introducing discretization errors due to the finite intended noise-level spacing $h=\lambda_{i+1}-\lambda_i$. Assuming initially that noise amplification is implemented perfectly, the true derivatives can be written as
\begin{align}
\label{eq:der_de}
\mathcal{A}^{(j)}_{\lambda_1}
=
\mathcal{A}^{[j]}_{\lambda_1}(M,h)
+
\mathcal{E}(M,h,j),
\end{align}
where $\mathcal{A}^{[j]}_{\lambda_1}(M,h)$ denotes the numerical derivative obtained from the available data points and $\mathcal{E}(M,h,j)$ represents the discretization error.

Using finite-difference formulas, these quantities can be expressed as
\begin{align}
\label{eq:num_der}
\mathcal{A}^{[j]}_{\lambda_1}(M,h)
&=
\sum_{i=1}^{M}
a_{ji}^{M}(h)\,
\mathcal{A}(\lambda_i),
\\
\mathcal{E}(M,h,j)
&=
\sum_{k=M}^{\infty}
b_{jk}^{M}(h)\,
\mathcal{A}^{(k)}_{\lambda_1}.
\end{align}

The coefficient matrices $a^{M}$ and $b^{M}$ depend on the number of noise scale factors $M$, the derivative order $j$, and the noise-level spacing $h$. Importantly, the rows of the finite-difference matrix satisfy
\begin{align}
\sum_{i=1}^{M} a_{ji}^{M}=0,
\end{align}
which guarantees that the numerical derivative vanishes when all data points are equal.

For the cases $M=2$ and $M=3$, the matrices used to compute the numerical derivatives are
\begin{align}
\label{eq:M2}
a^{M=2}(h) =
\begin{bmatrix}
-h^{-1} & h^{-1}
\end{bmatrix},
\end{align}
and
\begin{align}
\label{eq:M3}
a^{M=3}(h) =
\begin{bmatrix}
-\frac{3}{2}h^{-1} & 2h^{-1} & -\frac{1}{2}h^{-1} \\
h^{-2} & -2h^{-2} & h^{-2}
\end{bmatrix}.
\end{align}

In practice, noise amplification may deviate from the intended uniform spacing $h$. Instead, the implemented noise-scale spacings can be represented by a vector $\mathbf{t}$ with elements $t_i=\lambda_{i+1}-\lambda_i$. In this case the numerical derivatives are computed using modified finite-difference coefficients,
\begin{align}
\label{eq:num_der_2}
\mathcal{A}^{[j]}_{\lambda_1}(M,\mathbf{t})
=
\sum_{i=1}^{M}
a_{ji}^{'M}(\mathbf{t})\,
\mathcal{A}(\lambda_i).
\end{align}

The modified coefficients satisfy the same normalization condition
\begin{align}
\sum_{i=1}^{M} a_{ji}^{'M}=0,
\end{align}
ensuring that the numerical derivatives vanish if all data points are equal. For $M=2$, the coefficients are obtained from Eq.~(\ref{eq:M2}) by replacing $h$ with $t_1$. For $M=3$, defining $\tilde{t}=t_2/t_1$, one finds
\begin{align}
a^{'M=3}(\mathbf{t}) =
\begin{bmatrix}
-c_1(\tilde{t}^2 +2\tilde{t})  & c_1(1+\tilde{t})^2 & -c_1 \\
c_2\tilde{t} & -c_2(1+\tilde{t}) & c_2
\end{bmatrix},
\end{align}
where $c_1=[(1+\tilde{t})^2t_1 - (t_1+t_2)]^{-1}$ and $c_2=2[t_2(t_1+t_2)]^{-1}$.

Substituting Eq.~(\ref{eq:der_de}) into the Taylor expansion and accounting for imperfect noise amplification yields the residual bias of the baseline estimator,
\begin{align}
\label{eq:re}
\mathcal{B}_{\mathrm{b\text{-}NRE}}
=
&\sum_{k=M}^{\infty}
\left[
\frac{(-\lambda_1)^k}{k!}
+
\sum_{j=1}^{M-1}
\frac{(-\lambda_1)^j}{j!}
b_{jk}^{M}(h)
\right]
\mathcal{A}^{(k)}_{\lambda_1}
\nonumber\\
&+
\sum_{j=1}^{M-1}
\frac{(-\lambda_1)^j}{j!}
\left[
\sum_{i=1}^{M}
\left(
a_{ji}^{'M}(\mathbf{t})
-
a_{ji}^{M}(h)
\right)
\mathcal{A}(\lambda_i)
\right].
\end{align}

The first line represents the combined contributions from truncation and discretization errors, while the second line arises from imperfect noise amplification.

The structure of Eq.~(\ref{eq:re}) also provides insight into the relation between the magnitude of the residual bias $|\mathcal{B}_{\mathrm{b\text{-}NRE}}|$ and the normalized dispersion $\mathcal{D}$ observed in the main text. Both the truncation and discretization contributions to the residual bias are governed by the hierarchy of derivatives $\{\mathcal{A}^{(k)}_{\lambda_1}\}$.

On the other hand, the variation of $\mathcal{A}(\lambda_i)$ across the sampled
noise scales can be expanded as
\begin{align}
\mathcal{A}(\lambda_i)-\overline{\mathcal{A}}
=
\sum_{k=1}^{\infty}
\frac{1}{k!}
\left[
(\lambda_i-\lambda_1)^k
-
\overline{(\lambda-\lambda_1)^k}
\right]
\mathcal{A}^{(k)}_{\lambda_1},
\end{align}
where $\overline{\mathcal{A}}=(1/M)\sum_{i=1}^{M}\mathcal{A}(\lambda_i)$ and
\[
\overline{(\lambda-\lambda_1)^k}
=
\frac{1}{M}\sum_{i=1}^{M}(\lambda_i-\lambda_1)^k .
\]
Using the triangle inequality ($|\sum_k x_k| \le \sum_k |x_k|$), this implies
\begin{align}
\mathrm{MAD}[\{\mathcal A(\lambda_i)\}]
\le
\sum_{k=1}^{\infty}
\mu_k(M,\{\lambda_i\})
\,\big|\mathcal A^{(k)}_{\lambda_1}\big|,
\end{align}
where
\begin{align}
\mu_k(M,\{\lambda_i\})
=
\frac{1}{M\,k!}
\sum_{i=1}^{M}
\left|
(\lambda_i-\lambda_1)^k
-
\overline{(\lambda-\lambda_1)^k}
\right|.
\end{align}
Here MAD denotes the mean absolute deviation over the sampled noise scales.

This shows that both the residual bias $\mathcal{B}_{\mathrm{b\text{-}NRE}}$ and the numerator of the normalized dispersion are controlled by the same derivative hierarchy. In particular, when the higher-order derivatives $\mathcal{A}^{(k)}_{\lambda_1}$ are suppressed, both the residual bias and the variation of $\mathcal{A}(\lambda_i)$ across noise scales become small. Conversely, when these derivatives are large, they simultaneously increase the magnitude of the baseline-estimator bias and the variation of the auxiliary estimator across the sampled noise scales.

Importantly, the sign of $\mathcal{B}_{\mathrm{b\text{-}NRE}}$ depends on the detailed combination of the higher-order derivative contributions appearing in Eq.~(\ref{eq:re}), so different realizations can exhibit either positive or negative bias. By contrast, the normalized dispersion is constructed from mean absolute deviations and is therefore sensitive primarily to the overall scale of the variation of $\mathcal{A}(\lambda_i)$ rather than to the sign of the bias. In this sense, $\mathcal{D}$ tracks the magnitude of the derivative hierarchy that controls Eq.~(\ref{eq:re}) without uniquely determining the sign of the resulting bias.

In addition, the second line of Eq.~(\ref{eq:re}) vanishes when $\mathcal{A}(\lambda_i)$ becomes independent of $\lambda_i$, since both $\sum_i a_{ji}^{M}$ and $\sum_i a_{ji}^{'M}$ are zero. Consequently, in regimes where the auxiliary estimator exhibits weak dependence on the noise scale, the normalized dispersion $\mathcal{D}$ becomes small while all contributions to the residual bias are simultaneously reduced.

The normalization by $\mathrm{MAD}[\{\langle \tilde{O}\rangle_t(\lambda_i)\}]$ is important in this context because the absolute variation of $\mathcal{A}(\lambda_i)$ across the sampled noise scales depends not only on the underlying derivative structure but also on the spacing of the sampled noise levels. In particular, if one were to use only $\mathrm{MAD}[\{\mathcal{A}(\lambda_i)\}]$, the resulting quantity could be made artificially small simply by choosing a smaller noise-level spacing $h$. Dividing by the corresponding dispersion of the target-circuit expectation values removes this trivial scale dependence to leading order and makes $\mathcal{D}$ a relative measure of the variation of the auxiliary estimator with respect to the variation of the original noisy observable. This provides a theoretical justification for the observed correlation between the magnitude of the residual bias $|\mathcal{B}_{\mathrm{b\text{-}NRE}}|$ and the normalized dispersion $\mathcal{D}$ within a fixed experimental setting.
\section{Optimal Control Parameter}\label{subsec:n_op}

As explained in Sec.~\ref{sec:NRE_workflow}, by adjusting the control parameter to its
optimal value we render $\mathcal{A}(\lambda_1)$ the baseline estimator for
$\langle O\rangle_t$. Given the linear dependence of $\mathcal{A}$ on the control
parameter as defined in Eq.~(\ref{eq:aux_O}), it is straightforward to show
that the summation term in Eq.~(\ref{eq:Taylor_sum}) vanishes when the control
parameter is tuned to

\begin{align}\label{eq:n_op}
n_{\mathrm{op}}
=
-\frac{\sum_{i}\mathcal{V}_i^{M}\mathcal{P}_1(\lambda_i)}
{\sum_{i}\mathcal{V}_i^{M}\mathcal{P}_2(\lambda_i)},
\end{align}

where the coefficients $\mathcal{V}_i^{M}$ are defined as

\begin{align}\label{eq:V}
\mathcal{V}_i^{M}
=
\sum_{j=1}^{M-1}
\frac{(-\lambda_1)^j}{j!}\,
a_{ji}^{M}(h).
\end{align}

\begin{figure*}[t]\centering\includegraphics[width=0.8\textwidth]{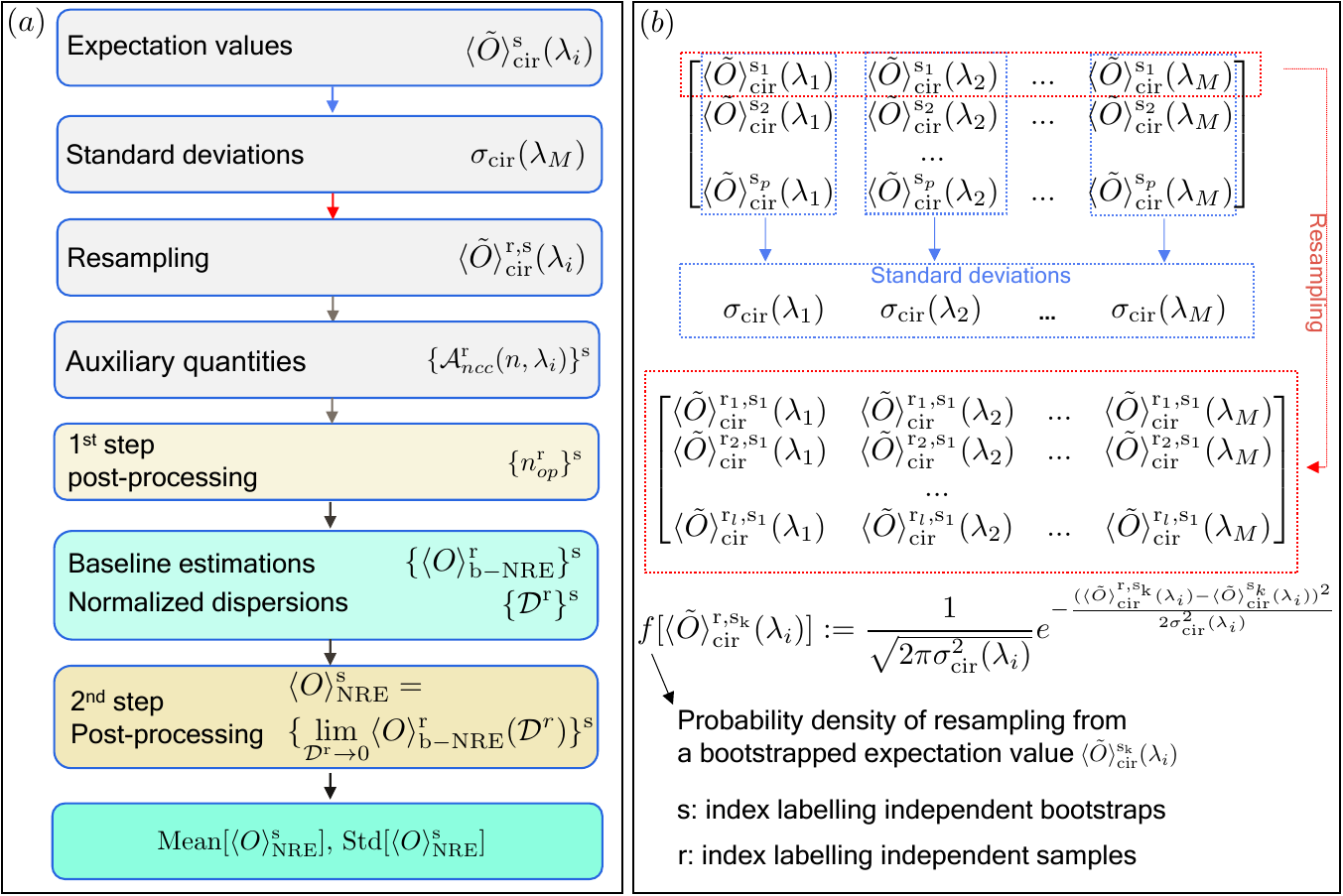}
\caption{
(a) Extended NRE framework used to compute the mean and standard deviation of the final estimator. Measurement counts are first bootstrapped to obtain bootstrapped expectation values at different noise scale factors. From these datasets the standard deviation of $\langle \tilde{O} \rangle_{\mathrm{cir}}$ (where ``cir'' denotes both the target circuit \textit{t} and the noise-canceling circuit \textit{ncc}) is estimated for each noise scale factor. Resampled expectation values are then drawn from normal distributions with means given by the bootstrapped expectation values and standard deviations obtained in the previous step. Each resampled dataset is processed through the NRE post-processing pipeline, producing a distribution of noise-robust estimates from which the mean and standard deviation of the final estimator are computed. (b) Schematic illustration of the resampling procedure applied to the set of bootstrapped expectation values.}
  \label{fig:NRE_extended_wf}
\end{figure*}
Since throughout this work we set $\lambda_1=1$, these coefficients simplify.
Using Eqs.~(\ref{eq:M2}) and (\ref{eq:M3}), one readily finds for $M=2$ and
$M=3$

\begin{align}\label{eq:V_M2}
\mathcal{V}^{M=2}
=
h^{-1}
\begin{bmatrix}
1 & -1
\end{bmatrix},
\end{align}

\begin{align}\label{eq:V_M3}
\mathcal{V}^{M=3}
=
h^{-1}
\begin{bmatrix}
\frac{3}{2} + \frac{1}{2}h^{-1} &
-2 - h^{-1} &
\frac{1}{2} + \frac{1}{2}h^{-1}
\end{bmatrix}.
\end{align}
The choice $n=n_{\mathrm{op}}$ therefore cancels the Taylor correction term in
Eq.~(\ref{eq:Taylor_sum}), rendering $\mathcal{A}(\lambda_1)$ the baseline
estimator for $\langle O\rangle_t$. In this sense, the optimal control parameter
tunes the contribution of the noise-canceling circuit such that the auxiliary
estimator becomes locally insensitive to noise scaling around $\lambda_1$.

\section{Extended NRE Procedure}
\label{subsec:extended_NRE}

As mentioned in the main text, one approach to determine the mean and standard
deviation of the final estimation in the NRE workflow
(Fig.~\ref{fig:NRE_Schematics}(a)) is to repeat the entire NRE workflow from
the stage of bootstrapped measurement counts a sufficiently large number of
times. While this approach is statistically valid, it requires a very large
number of bootstrap realizations (e.g., tens of thousands) for each
experimental dataset, making it computationally expensive on a classical
computer.

To mitigate this overhead while maintaining statistical accuracy, we adopt a
two-stage procedure illustrated in Fig.~\ref{fig:NRE_extended_wf}. First, we
perform bootstrapping on the measurement counts to obtain bootstrapped
expectation values
\(
\langle \tilde{O}_{\mathrm{cir}}^{\,s}(\lambda_i) \rangle
\)
for each noise scale factor $\lambda_i$. From these bootstrapped datasets we
estimate the corresponding standard deviations
\(
\sigma_{\mathrm{cir}}(\lambda_i)
\)
of the expectation values.

In the second stage, we generate a large number of resampled expectation
values
\(
\langle \tilde{O}_{\mathrm{cir}}^{\,r,s}(\lambda_i) \rangle
\)
from a normal distribution with

\begin{itemize}
\item mean given by the bootstrapped expectation value
\(
\langle \tilde{O}_{\mathrm{cir}}^{\,s}(\lambda_i) \rangle
\),
\item standard deviation given by
\(
\sigma_{\mathrm{cir}}(\lambda_i)
\).
\end{itemize}

Here $s$ labels independent bootstrap realizations and $r$ labels independent
resampling instances drawn from the corresponding normal distribution. The
resampled datasets therefore emulate repeated experimental realizations
consistent with the statistical uncertainty obtained from the bootstrapped
counts.

For each resampled dataset we apply the NRE post-processing layers,
producing a distribution of final NRE estimates
\(
\langle O \rangle_{\mathrm{NRE}}^{\,r,s}.
\)
The mean and standard deviation of the final estimator are then obtained from
this distribution.

Figure~\ref{fig:NRE_extended_wf}(a) illustrates the extended NRE workflow and
how the statistical uncertainty of the final estimate is propagated through
the NRE post-processing steps. Figure~\ref{fig:NRE_extended_wf}(b) provides a
schematic example of the resampling procedure applied to the set of
bootstrapped expectation values.

\section{Construction of Noise-Canceling Circuits}
\label{sec:ap_ncc}

The noise-canceling circuit (\textit{ncc}) is constructed from the target circuit while
preserving the circuit structure, including the number of gates, their
connectivity, and the order in which they appear in the circuit. To achieve
this, we first transpile the target circuit into a gate set consisting of the
two-qubit entangling gate $\mathrm{CZ}$ and single-qubit rotations
$R_X(\theta)$, $R_Y(\theta)$, and $R_Z(\theta)$. In this decomposition, the
only non-Clifford operations are the single-qubit rotations with arbitrary
angle $\theta$.

To construct the noise-canceling circuit, we create a copy of the transpiled
target circuit and replace each non-Clifford single-qubit rotation
$R_\alpha(\theta)$ ($\alpha \in \{X,Y,Z\}$) with a Clifford rotation
$R_\alpha(\theta_c)$, where the rotation angle $\theta_c$ is chosen from the
set $\{0,\pi/2,\pi,3\pi/2\}$. The angle $\theta_c$ is selected such that the
resulting Clifford unitary minimizes the Frobenius norm of the difference
between the Clifford operation and the corresponding non-Clifford gate in the
target circuit.

After performing these replacements, both the target circuit and the
noise-canceling circuit are transpiled to the native gate set of the backend.
For the experiments on the IQM device used in this work, the native gate set
consists of the two-qubit gate $\mathrm{CZ}$ and the single-qubit rotation
$r(\theta,\phi)$. The transpilation is carried out using settings that ensure
the structural equivalence of the two circuits is preserved, such that both
circuits experience comparable hardware-level noise processes.

\section{Different Regression Strategies}
\label{sec:ap_regression_strategies}

As discussed in Sec.~\ref{sec:TFIM_implementation}, the results presented in the main text are obtained using a weighted linear regression in which each baseline estimate $\langle O\rangle_{\mathrm{b\text{-}NRE}}^{s}$ is assigned a weight inversely proportional to its normalized dispersion,
\begin{align}
w_s = \frac{1}{\mathcal{D}_s}.
\end{align}
This weighting scheme prioritizes baseline realizations with smaller normalized dispersion, which are expected to lead to smaller residual bias magnitude.

\begin{figure}[t]
\centering
\includegraphics[width=1.1\linewidth]{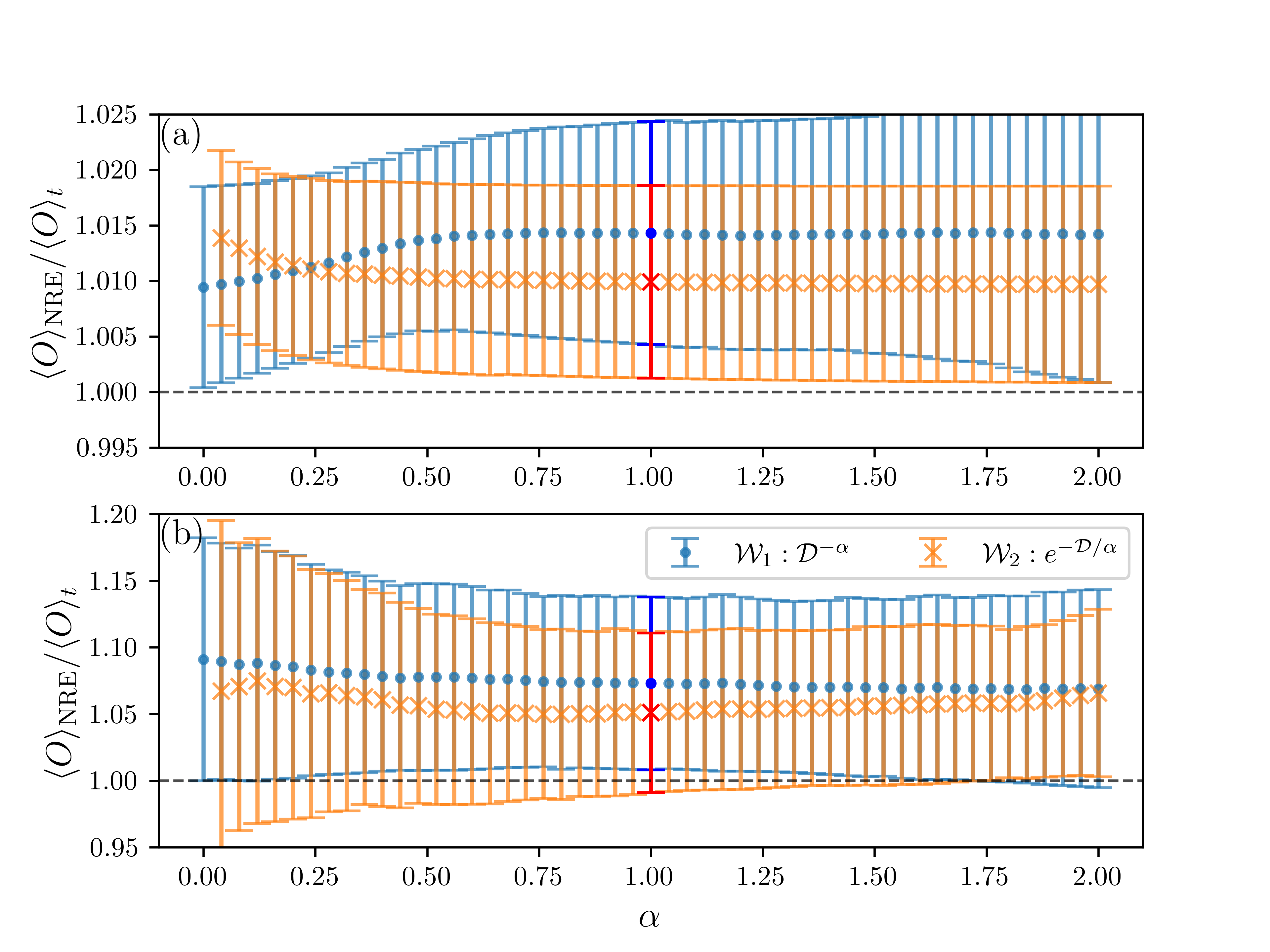}
\caption{Robustness of the NRE estimator with respect to the regression weighting strategy.
(a) TFIM benchmark corresponding to the experiment shown in Fig.~\ref{fig:NRE_on_Garnet}(a). (b) $\mathrm{H}_4$ molecular benchmark corresponding to Fig.~\ref{fig:NRE_Benchmarks_H4}(a).
Blue circles correspond to the power-law weighting $\mathcal{W}_1 = \mathcal{D}^{-\alpha}$, while orange crosses correspond to the exponential weighting $\mathcal{W}_2 = \exp(-\mathcal{D}/\alpha)$.
Error bars are obtained from bootstrap resampling. Data points with darker colors indicate $\alpha=1$, corresponding to the weighting used in the main text. The horizontal dashed line marks the ideal value $\langle O\rangle_{\mathrm{NRE}}/\langle O\rangle_t = 1$.
Across a broad range of $\alpha$, the final NRE estimate remains stable, demonstrating that the regression procedure does not require fine tuning.}
\label{fig:weight_benchmarks}
\end{figure}
In this appendix we investigate the sensitivity of the final NRE estimate to the precise form of the regression weighting. In particular, we consider two families of weight functions parameterized by a real parameter $\alpha$,
\begin{align}
\mathcal{W}_1(\alpha) = \frac{1}{\mathcal{D}_s^{\alpha}},
\end{align}
and
\begin{align}
\mathcal{W}_2(\alpha) = \exp\!\left(-\frac{\mathcal{D}_s}{\alpha}\right).
\end{align}

The weighting scheme used throughout the main text corresponds to the choice $\mathcal{W}_1$ with $\alpha = 1$. Varying $\alpha$ therefore allows us to systematically control how strongly baseline realizations with larger normalized dispersion are suppressed in the regression procedure.

Figure~\ref{fig:weight_benchmarks} shows the resulting NRE estimates obtained using these weighting strategies. In Fig.~\ref{fig:weight_benchmarks}(a) we analyze the TFIM experiment presented earlier in Fig.~\ref{fig:NRE_on_Garnet}(a), while Fig.~\ref{fig:weight_benchmarks}(b) shows the corresponding analysis for the $\mathrm{H}_4$ benchmark discussed in Fig.~\ref{fig:NRE_Benchmarks_H4}(a).

In both cases the final NRE estimate exhibits only a weak dependence on the precise form of the regression weights across a broad range of $\alpha$. In particular, a wide plateau is observed around $\alpha \approx 1$, indicating that the weighting choice used in the main text lies within a stable region and does not require fine tuning. This behavior reflects the fact that the dispersion–bias correlation exploited by the NRE protocol is sufficiently strong that moderate variations of the regression strategy do not significantly affect the resulting estimator.

We also observe that the exponential weighting $\mathcal{W}_2$ can lead to a slight improvement in the final estimate compared with the power-law weighting used in the main text. However, the differences remain small, suggesting that the overall performance of NRE does not critically depend on the detailed choice of regression weighting.

\bibliography{papers}

\end{document}